\documentclass[aps,floatfix,superscriptaddress,notitlepage]{revtex4-1}
\usepackage{amsmath,amssymb,amsfonts,graphics,graphicx,dcolumn,bm,enumerate}
\usepackage{comment,natbib}
\usepackage[ddmmyyyy,hhmmss]{datetime}
\usepackage{multirow,color}

\newcommand{\sinp}
{\affiliation{Condensed Matter Physics Division, 
Saha Institute of Nuclear Physics, 1/AF Bidhannagar, Kolkata 700064, India.}}
\newcommand{\isi}
{\affiliation{Economic Research Unit, Indian Statistical Institute, 203 B. T. 
Road, Kolkata 700108, India.}}
\newcommand{\aalto}
{\affiliation{Department of Computer Science, Aalto University School of 
Science, P.O. Box 15400, FI-00076 AALTO, Finland.}}

\begin{document}

\title{Universality of citation distributions for academic institutions and 
journals}

\author{Arnab Chatterjee}%
\email[Email: ]{arnabchat@gmail.com}
\sinp
\author{Asim Ghosh}
\email[Email: ]{asim.ghosh@aalto.fi}
\sinp \aalto
\author{Bikas K. Chakrabarti}
\email[Email: ]{bikask.chakrabarti@saha.ac.in}
\sinp \isi

\begin{abstract}
 Citations measure the importance of a publication, and may serve as a proxy
for its popularity and quality of its contents.
Here we study the distributions of citations to publications from 
individual academic institutions for a single year. 
The average number of citations have large variations between different 
institutions across the world,
but the probability distributions of citations for individual institutions
can be rescaled to a common form by scaling the citations by the average number 
of 
citations for that institution.
We find this feature seem to be universal for a broad selection of 
institutions 
irrespective of the average number of citations per article.
A similar analysis for citations
to publications 
in a particular journal in a single year 
reveals similar results.
We find high absolute inequality for both these sets, Gini coefficients
being around $0.66$ and $0.58$ for institutions and journals respectively.
We also find that the top $25$\% of the articles hold about $75$\% of the total 
citations for institutions and the top $29$\% of the articles hold about $71$\% 
of the total citations for journals.
\end{abstract}

\maketitle 
\section*{Introduction}
Statistical physics tells us that systems of many interacting dynamical units 
collectively exhibit a behavior which is
determined by only a few basic dynamical features of the individual units and 
of 
the embedding dimension but independent of all other details.
This  feature which is specific to critical phenomena, like in continuous phase 
transitions, is known as 
universality~\cite{stanley1971introduction}. 
There is enough empirical evidence that a number of social phenomena are 
characterized by simple emergent behavior out of the
interactions of many individuals. 
In recent  years, a growing community of researchers have been analyzing 
large-scale 
social dynamics to uncover universal patterns and also trying to propose simple 
microscopic models to describe them, 
similar to the minimalistic models used in statistical physics.
These studies have revealed interesting patterns and behaviors in social 
systems, e.g.,
in 
elections~\cite{fortunato2007scaling,chatterjee2013universality,
mantovani2011scaling}, 
growth in population~\cite{rozenfeld2008laws} and 
economy~\cite{stanley1996scaling},
income and wealth distributions \cite{chakrabarti2013econophysics}, 
financial markets~\cite{mantegna2000introduction}, 
languages~\cite{petersen2012statistical}, etc.
(see Refs.~\cite{Castellano:2009,Sen:2013} for reviews).

Academic publications (papers, books etc.) form an unique social system 
consisting
of individual publications as entities, containing bibliographic reference to 
other older publications,
and this is commonly termed as \textit{citation}. 
The number of citations is a measure of the importance of a publication, and 
serve as a proxy
for the popularity and quality of a publication.
There has already been a plethora of empirical studies on citation 
data~\cite{Sen:2013}, specifically on 
citation 
distributions~\cite{Shockley:1957,Laherrere:1998,Redner:1998,Radicchi:2008} of 
articles,  
time evolution of probability distribution of 
citation~\cite{Rousseau:1994,Egghe:2000,Burrell:2002}, 
citations for individuals~\cite{Petersen:2011} and even their 
dynamics~\cite{Eom:2011},
and the modeling efforts on the growth and structure of citation networks have 
produced a huge body literature
in network science concerning scale-free 
networks~\cite{Price:1965,barabasi1999emergence,caldarelli2007scale},
and long-time scientific impact~\cite{wang2013quantifying}.

The bibliometric tool of citation analysis is becoming increasingly popular
for evaluating the performance of individuals, research groups, institutions as 
well as countries,
the outcomes of which are becoming important in case of offering grants and 
awards, 
academic promotions and ranking, as well as jobs in academia, industry and 
otherwise.
Since citations serve as a crucial measure for the importance and impact of a 
research
publication, its precise analysis is extremely important.
Annual citations and impact factor of journals are of key interest, primarily
from the point of view of journals themselves, and secondarily from the 
perspective of authors
who publish their papers in them. Wide distributions of both annual citations 
and impact factors
are quite well studied~\cite{Popescu:2003,Mansilla:2007,khaleque2014evolution}.
It is quite usual to find that some publications do better than others
due to the inherent heterogeneity in the quality of their content, the gross 
attention 
on the field of research, the relevance to future work and so on. Thus different 
publications 
gather citations in time at different rates and result in a broad distribution 
of citations.
In 1957, Shockley~\cite{Shockley:1957} claimed that the scientific publication 
rate
is dictated by a lognormal distribution, while a later evidence based on 
analysis of records for
highly cited physicists claim that the citation distribution of individual 
authors follow a stretched
exponential~\cite{Laherrere:1998}. However, an analysis of data from ISI claims 
that the tail of the citation
distribution of individual publications decays as a power law with an exponent 
close to $3$~\cite{Redner:1998},
while a rigorous analysis of 110 years of data from Physical Review concluded 
that most part of
the citation distribution fits remarkably well to a 
lognormal~\cite{Redner:2005}.
The present consensus lies with the fact that while most part of the 
distribution does
fit to a lognormal, the extreme tail fits to a power 
law~\cite{peterson2010nonuniversal}.

It has been shown earlier~\cite{Radicchi:2008} that the 
distribution of citations $c$
to papers within a discipline has a broad distribution, which is universal 
across broad scientific
disciplines, using a relative indicator $c_f=c/\langle c \rangle$, where 
$\langle c \rangle$ is the 
average citation within a discipline. However, it has also been shown later that 
this universality
is not absolutely guaranteed~\cite{waltman2012universality}.
Subsequent work on citations~\cite{radicchi2011rescaling,radicchi2012reverse} 
and impact factors~\cite{kaur2013universality} has revealed interesting 
patterns of universality, some alternative methods have been 
proposed~\cite{li2013quantitative} and there are also interesting work on
citation biases~\cite{radicchi2012reverse}.
Some studies~\cite{albarran2011references,albarran2011skewness} also report on 
the possible lack of universality in the citation distribution at the level of 
articles.
A rigorous and detailed study on the citation distributions of papers 
published in 2005-2008 for 500 institutions~\cite{private} reveals that 
using the analysis Ref.~\cite{Radicchi:2008}, universality condition is not 
fully satisfied, but the distributions are found to be very similar.
There have also been studies at the level of countries in the same 
direction~\cite{albarran2013differences}.

In this article, we focus on citations received by individual (i) academic 
institutions and (ii) academic journals.
We perform the analysis primarily for all articles and reviews, as well as all 
citable documents. 
While institutions can vary in their quality of scientific output measurable in 
terms of total number
of publications, total citations etc., here we show for the first time  that 
irrespective 
of the institution's scientific productivity, ranking and research impact, the 
probability $P(c)$
that the number of citations $c$
received by a publication is a broad distribution with an universal functional 
form. 
In fact, using a relative indicator 
$c_f=c/\langle c \rangle$, where $\langle c \rangle$ is the average number of 
citations to articles
published by an institution in a certain year, we show that the effective 
probability distribution function
that an article has $c$ citations has the same mathematical form. We present 
evidence for the fact that this holds roughly across time for most 
institutions 
irrespective of the scientific productivity of the institution considered.
When we carry out a similar analysis on journals,
we find similar results.
The scaled distributions fit to a lognormal distribution
for most of their range. Again, we find that these features roughly hold across 
time and across journals within the same class.
The largest citations for academic institutions as well as the journals
seem to fit well to a power law.
We also present evidence that each of these sampled groups -- institutions, and 
journals are distinct with the absolute measure of inequality as 
computed from their distribution functions, with high absolute inequality for 
both these sets, the Gini coefficients being around $0.66$ and $0.58$ for 
institutions and journals respectively.
We also find that the top $25$\% of the articles fetch about $75$\% of the 
total citations for institutions and the top $29$\% of the articles fetch about
$71$\% of the total citations for journals.

\section*{Methods}

\subsubsection*{Data}
We collected data from 42 academic institutions across the world. Institutions 
were selected such that
they produce considerable amount of papers (typically 200 or more) so that 
reasonable statistics could be obtained. However, there were exceptions for 
certain years for particular institutions.
All papers published with at least one author
with the institution mentioned as affiliation were collected. This was done for 
4 years -- 1980, 1990, 2000, 2010.
We also selected 30 popular academic journals across physics, chemistry, biology 
and medicine.
However, for some journals, only 3 years of data could be collected, since they 
were launched after 1980.
The citable papers considered in this study are articles and reviews, although 
we compare the results with the same analysis done on all citable documents.

\section{Results}
We study the data of number of citations to publications from different years,
from ISI Web of Science~\cite{ISI} for several  
(i) academic institutions (research institutes and universities) and (ii) 
popular journals.
It is to be noted that citations to individual publications arrive from any 
publication indexed in ISI Web of Science
and does not mean only internal citations within the journal in which it is 
published.
We analyzed data of science publications from 42 academic institutions and 30 
popular journals.
We recorded the data for the number of papers published, the total number of 
citations to each of the
publications, for a few years (1980, 1990, 2000, 2010 for most cases). 
Since citations grow with time, we have studied publications which are at least 
4 years old (from 2010) or more (1980, 1990, 2000)
to rule out any role of transients. We also collect data from academic 
institutions and journals which have a comparatively
large number of publications, so as to produce good statistics, and minimize the 
effects of aberration that can result
from fluctuations of the quantities measured from small data sets.

\subsubsection{Citations for academic institutions}
\begin{figure*}[h]
\includegraphics[width=15.5cm]{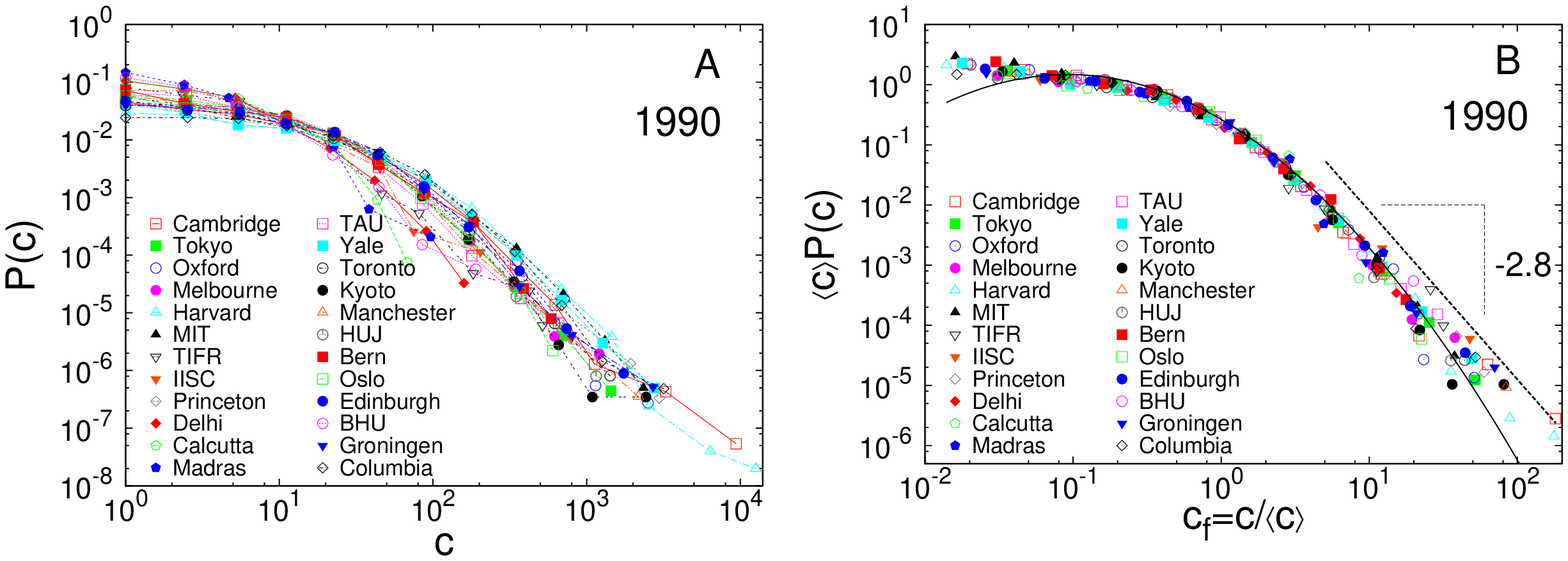}
\caption{\textbf{Probability distribution of citation for academic institutions 
for 1990 (unscaled and rescaled).}\\
(A) Probability distribution $P(c)$ of citations $c$ to publications from 1990 
for several academic institutions. 
(B) The same data rescaled by average number of citations $\langle c \rangle$.
The data for different institutions seem to follow the same scaling
function. It fits very well to a lognormal function for most of its range, 
with $\mu = -0.73 \pm 0.02$, $\sigma = 1.29 \pm 0.02$.
The largest citations do not follow the lognormal behavior, and seem to follow 
a 
power law: $c^{-\alpha}$, with $\alpha = 2.8 \pm 0.2$.
}
\label{fig:inst_all_1990}
\end{figure*}

\begin{figure*}[h]
\includegraphics[width=15.5cm]{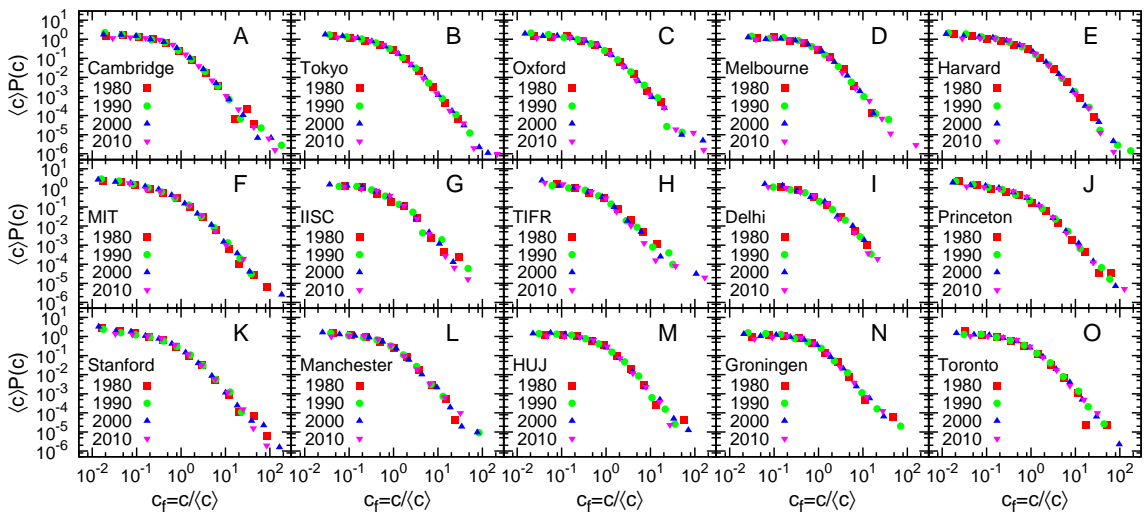}
\caption{\textbf{Rescaled probability distributions of citation for several 
academic institutions for different years.}\\
Probability distribution $P(c)$ of citations $c$ rescaled by average number of 
citations $\langle c \rangle$
 to publications from 4 different years (1980, 1990, 2000, 2010) for several 
academic institutions.
For any institution, the data for different years seem to follow the same 
distribution.
}
\label{fig:inst_years}
\end{figure*}

We collected citation data until date for all articles and reviews from a 
particular year (e.g. 1980, 1990, 2000, 2010). 
For each year, the probability distribution $P(c)$ of citations $c$ for an 
academic institution was observed to be broad.
For instance, Fig.~\ref{fig:inst_all_1990}A shows the plot of $P(c)$ vs. $c$ for 
various institutions 
for publications from 1990. 
We rescaled the absolute value of citation for each year by the average number 
of citations per 
publication $\langle c \rangle$,
and plotted this quantity $c_f=c/\langle c \rangle$ against the adjusted 
probability $\langle c \rangle P(c)$ (Fig.~\ref{fig:inst_all_1990}B)
(see similar plots for 1980 and 2000 in Fig.~S1 of SI).
We remarkably find that the distributions collapse into an universal curve 
irrespective
of the wide variation in the academic output of the different institutions.
The scaling collapse is good for more than 3 decades of data and over 5 orders 
of magnitude.
The average number of papers, total citations and the average number of 
citations per publication 
are shown in Table~S3. 
The rescaled curves fit well to a lognormal
\begin{equation}
 F(x) = \frac{1}{x \sigma \sqrt{2\pi}} \exp \left[-\frac{(\log x - \mu)^2}{2 
\sigma^2}\right]
 \label{eq:ln}
\end{equation}
with $\mu =-0.73 \pm 0.02$ with $\sigma = 1.29 \pm 0.02$, for a considerable 
range of the distribution. However, if one fits a lognormal distribution to 
individual sets, the range of parameters are quite narrow, $\mu$ lies in the 
range $-1.2$ to $-0.6$, while $\sigma$ lies in the range $1.0$ to $1.6$.
The fitting were performed using a least square fitting routine.
For lowest values of the abscissa, seems to follow $\langle c \rangle P(c) \to 
const$ or slowly growing, as $c \to 0$.
However, the largest citations deviate from the lognormal fit and are better
described according to $P(c) \sim c^{-\alpha}$, with $\alpha = 2.8 \pm 0.2$ 
(see SI Table~S6 for exponents for other years).
The power law exponent has been estimated using the maximum likelihood estimate 
method (MLE)~\cite{clauset:2009}.
In order to investigate if the distributions $P(c)$ for different institutes 
vary with time, 
we plot the same for each institution for several years. 
The rescaled plots show scaling collapse indicating that although the 
average citations
vary over years, the form of the distribution function remain roughly 
invariant, when scaled with the average number of citations. 
Fig.~\ref{fig:inst_years} shows the plot for 1990.
To check if this also holds for time-aggregated data, we collected citations for 
all papers published during the period
2001-2005 for the same set of institutions, and repeated the above analysis 
(see SI. Fig.~S2).


\subsubsection{Citations for journals}
\begin{figure*}[h]
\includegraphics[width=15.5cm]{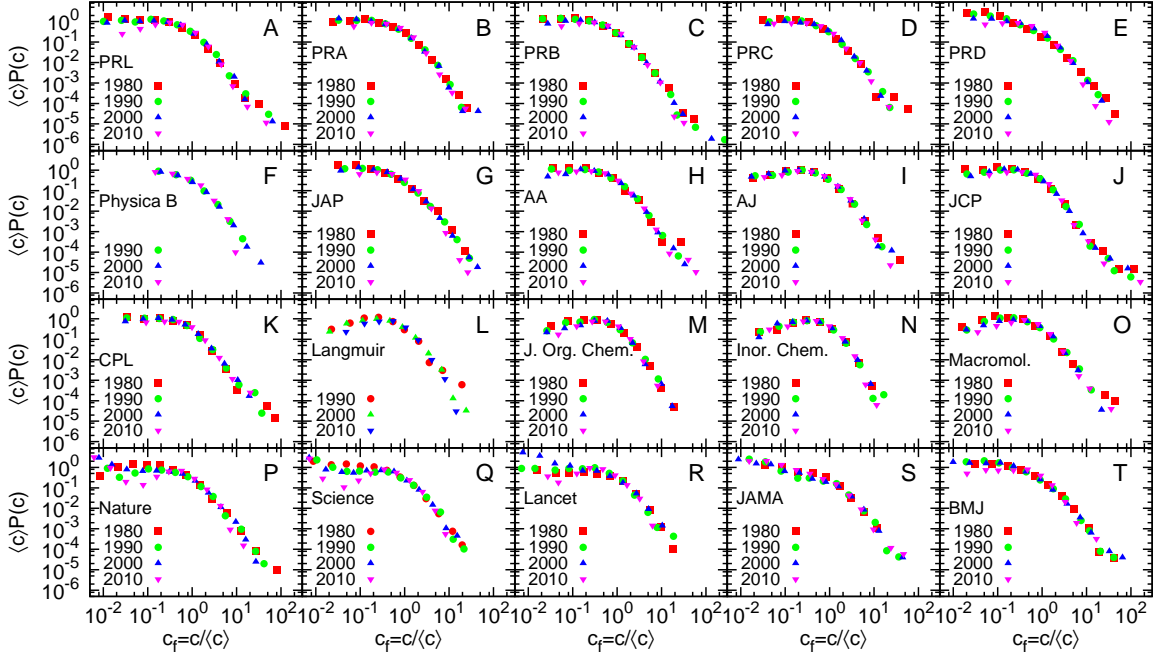}
\caption{\textbf{Rescaled probability distributions of citation for academic 
journals for different years.}\\
Probability distribution $P(c)$ of citations $c$ rescaled by average number of 
citations $\langle c \rangle$
to publications from from 4 different years (1980, 1990, 2000, 2010)  for 
several academic journal. 
For any journal, the data for different years seem to follow the same 
distribution.}
\label{fig:joun_all}
\end{figure*}

\begin{figure*}[h]
\includegraphics[width=10.0cm]{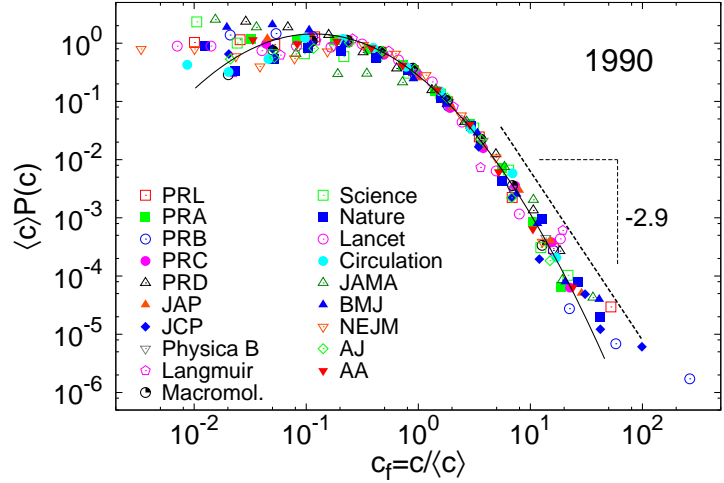}
\caption{\textbf{Rescaled probability distributions of citation for several 
journals for 1990.}\\
Probability distribution $P(c)$ of citations $c$ rescaled by average number of 
citations $\langle c \rangle$
to publications from 1990 for several academic journals.
The scaled distribution fucntion fits  to a lognormal 
function with $\mu = -0.75 \pm 0.02$,  $\sigma = 1.18 \pm 0.02$, 
while $\langle c \rangle P(c) \to const.$ as $c/\langle c \rangle \to 0$
for the lower range of $c$.
The largest citations fit well to a power law: $c^{-\alpha}$, with $\alpha = 
2.9 \pm 0.3$.
}
\label{fig:jour_class_1990}
\end{figure*}

We collected citation data until date for all articles and reviews in 
individual journals
for several years (e.g. 1980, 1990, 2000, 2010 etc.). 
For each year, the probability distribution $P(c)$ of citations $c$ was again 
observed to be broad. 
As in the case of institutions, we plotted $c_f=c/\langle c \rangle$ against 
the 
adjusted probability $\langle c \rangle P(c)$
(Fig.~\ref{fig:joun_all}). 
For a particular journal, it is observed that the curves follow similar 
distributions over years
although the average number of papers, total citations and hence the average 
number of citations vary
(See Table~S4 in SI for details).
Further, we plot the same quantity for a particular year for different journals 
(see Fig.~\ref{fig:jour_class_1990} for 1990),
and find that the curves roughly
collapse into an single curve irrespective
of the wide variation in the output of the different journals.
The bulk of the rescaled distribution fits well to a lognormal form 
with $\mu = -0.75 \pm 0.02$ and $\sigma =1.18 \pm 0.02$, as 
was observed 
in case of institutions, while the largest citations fit better to a power law
$P(c) \sim c^{-\alpha}$, with $\alpha \approx 2.9 \pm 0.3$ 
(see SI Table~S6 for exponents for other years).
To check if this also holds for time-aggregated data, we collected citations 
for all papers published during the period
2001-2005 for the same set of journals and repeated the analysis (see SI. 
Fig.~S4).

However, we observed that if we consider all citable documents, 
two distinct classes of journals emerge
according to the shape of the distributions to which the curves 
collapse.
The first group is a \textit{General} class, 
for which most of the distribution fits well to a lognormal function
even quite well for the lowest values of the abscissa.
This is similar to what is observed for all journals if we consider only 
articles and reviews.
The other group, which we call the \textit{Elite} class  (SI Fig.~S5)
is also broadly distributed but has a distinct and faster monotonic decay 
compared to
the \textit{General} class, 
where $\langle c \rangle P(c) \sim (c/\langle c \rangle)^{-b}$,
i.e., $P(c) \sim c^{-b}/\langle c \rangle^{(1-b)}$  with $b \simeq 1$.
This divergence at the lowest values of citations also indicate that the 
\textit{Elite} journals have 
a larger proportion of publications with less number of citations although their 
average number 
of citations $\langle c \rangle$ is larger than those for the general class.
However, for both the above classes, the largest citations still follow a power 
law $P(c) \sim c^{-\alpha}$, with $\alpha \approx 2.8 \pm 0.4$
for the \textit{General} class and $\alpha \approx 2.7 \pm 0.6$ 
for the \textit{Elite} class.
We reason for such a behavior in the \textit{Elite} journal class is because of 
a large fraction $f_0$ of uncited documents. If we consider only articles and 
reviews, $f_0$ is usually 2-10\%. Considering all citable documents, this 
fraction does not change appreciably for the \textit{Elite} class of journals, 
and can be anything in the range 25-80\% (see values of $f_0$ in SI Table 
S5), and are primarily in the category of news, correspondence, editorials etc. 
Such documents in the \textit{General} class is either absent or are very few.
We are able to find at least 7 journals (see SI Fig.~S5 and Table~S5)
in the \textit{Elite} class while most of others belong to the \textit{General} 
class.

The power law tail in all distribution 
suggests that the mechanism behind the popularity of the very highly cited 
papers is a `rich gets richer'
phenomena~\cite{Merton:1968,Price:1976,barabasi1999emergence} (see Fig.~S3 of SI 
for 1980 and 2000).

%

\subsubsection{Justification for using $c_f=c/\langle c \rangle$}
\begin{figure*}[h]
\includegraphics[width=15.5cm]{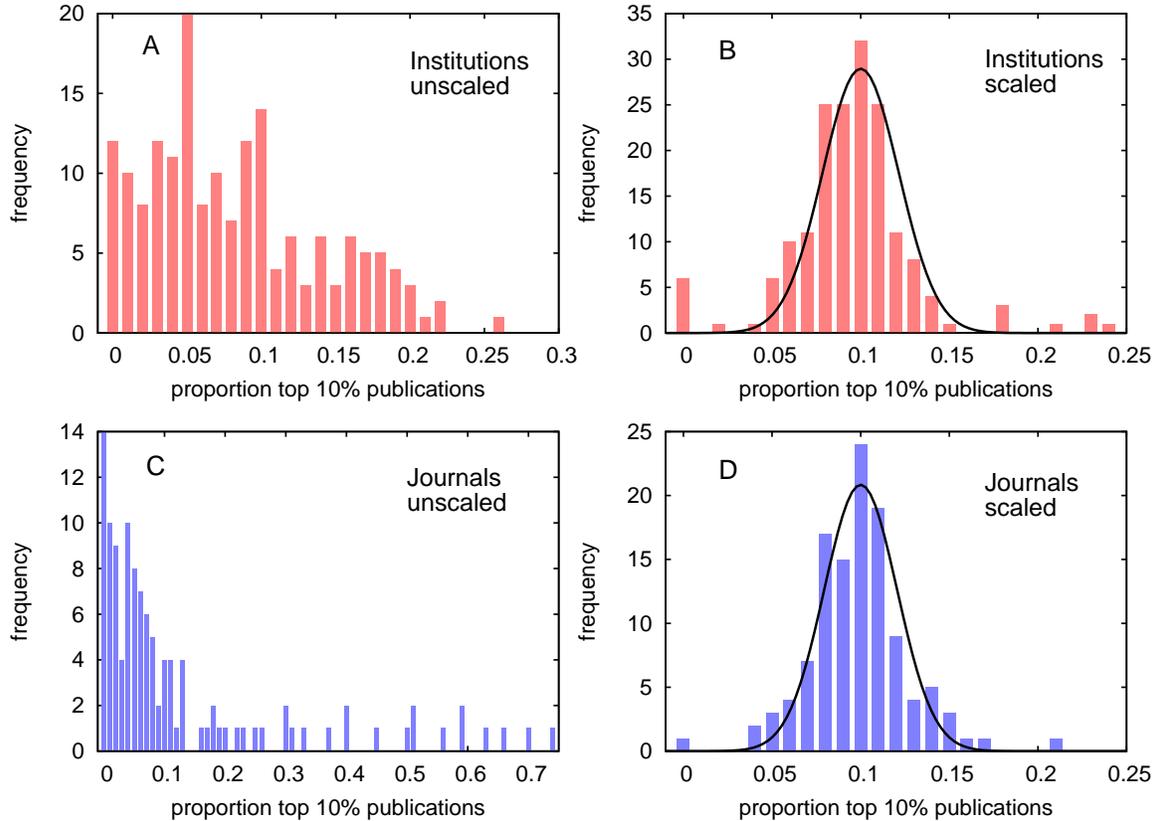}
\caption{\textbf{Percentage of publications of each institution that appear in 
the top 10\% of the global rank}\\
Histograms for the  
percentage of publications of each institution that appear in the top 10\% of 
the global rank, computed from the (A) unscaled and (B) scaled data.
Same for the  
percentage of publications of each journal that appear in the top 10\% of 
the global rank, computed from the (C) unscaled and (D) scaled data.
A normal distribution fit to the scaled data gives $\sigma_z =2.15 
\pm 0.08$ for institutions and $\sigma_z =2.05 \pm 
0.08$ for journals.
}
\label{fig:top10}
\end{figure*}

Following Ref.~\cite{Radicchi:2008}, we rank all articles belonging to 
different institutions according  to $c$ and  $c_f$. 
We then compute the percentage of publications of each institution that 
appear in the top $z$\% of the global rank. 
The percentage for each should be around $z$\% with small 
fluctuations if the ranking is good enough.
The same is performed for journals.
When ranking is done according to 
unnormalized citations $c$ then the frequency distribution of $z$\% of papers 
is wide. 
However, if the ranking is done according to normalized citations $c_f$,
then the frequency distribution is much narrow.
For example, we show the results for institutions and journals in 
Fig.~\ref{fig:top10} if $z=10$\%.
Assuming that articles are uniformly distributed on the rank axis, the 
expected average bin height must be z\% with a standard deviation given by
\begin{equation}
 \sigma_z =\sqrt{\frac{z(100-z)}{N} \sum_{i=1}^N \frac{1}{N_i}}.
\end{equation}
where $N$ is the number of entries (institutions or journals) and $N_i$ is the 
number of papers for the $i$-th institution or journal.
For institutions, when the ranking is done according to $c_f$ we observe that 
the theoretically calculated value (from above equation) of $\sigma_z$ is 
$1.25$ compared to $2.15 \pm 0.08$ as computed directly from the fitting, 
while if the ranking was done according to $c$, $\sigma_z$ is $6.29$.
Similarly for journals, $\sigma_z$ computed from the above equation is $1.04$ 
compared to $2.05 \pm 0.08$ as computed from the fitting, while if
the ranking was done according to $c$, $\sigma_z$ is $17.73$.
This indicates that $c_f$ is indeed an unbiased indicator, as seen 
earlier~\cite{Radicchi:2008,waltman2012universality}.

\subsubsection{Measuring inequality}
We calculate absolute measures of inequality like the commonly used Gini 
index~\cite{gini1921measurement}
as well as the  $k$-index~\cite{ghosh2014inequality,inoue2014measuring} which 
tells us that the top cited $1-k$ fraction of papers
have $k$ fraction of citations, and we report in SI Tables.~S3, S4, S5.
For academic institutions, Gini index $g = 0.67 \pm 0.10$  and $k= 0.75 \pm 
0.04$, which means around $75\%$ citations
come from the top $25\%$ papers.
For journals, $g=0.58 \pm 0.15$, $k= 0.71 \pm 0.08$
which means about $71\%$ citations
come from the top $29\%$ papers.

\begin{figure*}[h]
\includegraphics[width=15.5cm]{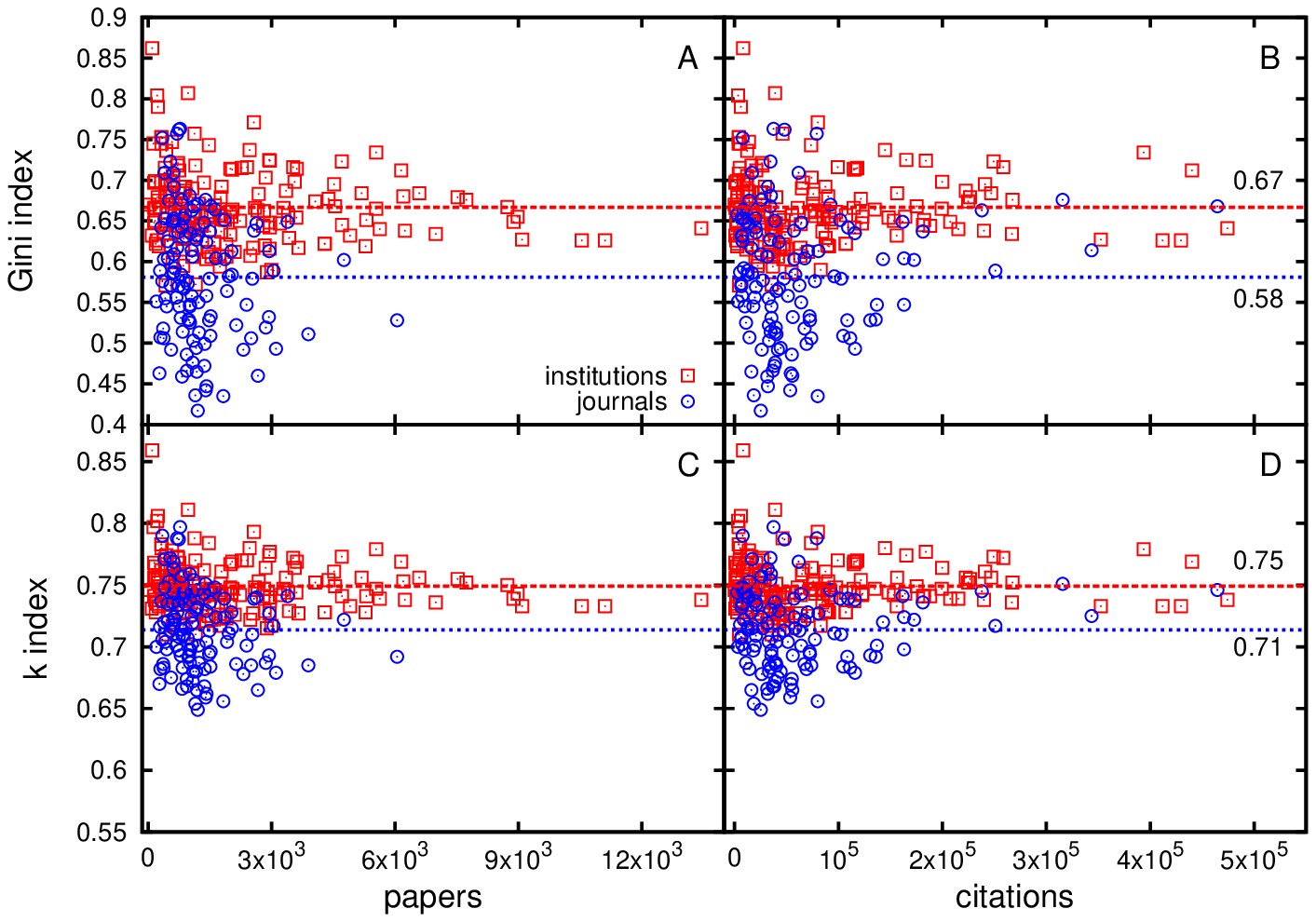}
\caption{\textbf{
Gini and $k$ indices with number of papers and 
citations}\\
Variation of Gini and $k$ indices  with number of papers and citations for 
academic institutions and journals.
For larger number of papers or citations, the values seem to fluctuate less or 
converge around the mean values $\bar{g}$ and $\bar{k}$ respectively.
For academic institutions, the values are $\bar{g} \approx 0.67$ for Gini and 
$\bar{k} \approx 0.75$, while for the journals, the values are $\bar{g} \approx 
0.58$ and $\bar{k} \approx 0.71$.
}
\label{fig:gk_conv}
\end{figure*}

We further note that Gini and $k$ indices
fluctuate less around respective mean values $\bar{g}$ and $\bar{k}$ as 
the number of articles and number of citations 
become large (Fig.~\ref{fig:gk_conv}).
For academic institutions, the values are $\bar{g} \approx 0.66$ for Gini and 
$\bar{k} \approx 0.75$. For journals, the values
are $\bar{g} \approx 0.58$ and $\bar{k} \approx 0.71$.

\section{Discussions}
In this article we analyze whether the 
citations to science publications from academic institutions (universities, 
research institutes etc.) as well as journals are distributed according
to some universal function when rescaled by the average number of citations.
For institutions, 
it seems to fit roughly
to a  log-normal function.
The largest citations, however, deviate from the lognormal fit, and follow a 
power law decay. 
This rough universality claim is an interesting 
feature, since for institutions, 
the quality of scientific output measurable in terms of the total number of 
publications, total citations etc. vary widely across the world as well as in 
time. Nevertheless, the way in which the number of papers with a certain number 
of citations is distributed is quite similar~\cite{private}, 
seems to be quite independent of the quality of 
production/output of the academic institution.
Although there has been claims that the form of the distribution of citations 
for different scientific 
disciplines are the same~\cite{Radicchi:2008}, 
albeit deviations~\cite{waltman2012universality}, 
it is also true that each discipline is characterized
by a typical average number of citations $\langle c \rangle_{d}$.
As a matter of fact, that different institutions have a varying strength of 
publication contribution
towards different disciplines makes the issue of obtaining a universal function 
for the resulting (effective)
distribution of citations (for the institution) quite nontrivial.
In other words,  different academic institutions have a variety in the strength 
of their academic output,
in terms of variation of representations across different disciplines and the 
amount of citations gathered.
This does not necessarily guarantee that the universality which has been 
already 
reported across
disciplines~\cite{Radicchi:2008} will still hold when one looks at data from 
different institutions,
rest aside the counter claims about  lack of universal 
character~\cite{waltman2012universality}
for citation distribution across distinct disciplines.
There are already critical studies on the citation distribution of 
universities~\cite{private} using larger data sets, which raises issues on the 
nature of universality.

We observe similar features for academic journals -- the bulk of the 
probability 
distribution fitting reasonably well to a lognormal while
the highest cited papers seem to fit well to a power law decay with a similar 
exponent ($2.7-3.0$).
We note that the exponents are consistently less than $3$, the exponent of the 
full citation distribution~\cite{Redner:1998},
which is due to the fact that our data are very small subsets, which fall short 
of catching the correct
statistical behavior of all of the highest cited papers.

Our results indicate that dividing citation counts by their average indeed 
helps to get closer to universal citation distributions. However, the results 
also indicate that, even after such a rescaling, $\sigma_z$ are substantially 
larger than the theoretical values -- $1.25$ compared to $2.15 \pm 0.08$ while 
it is $6.29$ for unscaled data  for 
institutions and $1.04$ compared to $2.05\pm 0.08$ while it is $17.73$ for 
unscaled data for journals.
This indicates that the universality is not 
very strong,  and holds only in an approximate sense.
Ref.~\cite{private} shows similar evidence for institutions, claiming the 
absence of universality but pointing out the similarity between the 
distributions. Another previous 
study~\cite{waltman2012universality} on different fields of science also 
reported that this universality claim does not hold very well for all fields.

We further note that the inequality in the distribution of citations of 
institutions and journals differ quantitatively. As the number of papers and 
citations increase, the absolute 
measures of inequality like Gini and $k$ indices seem to converge to different 
values for the above two sets. 
The values of Gini index are $0.66$ and $0.58$ for institutions and journals 
respectively. 
The $k$ index values suggest that the top $25$\% of the articles hold about 
$75$\% of the total 
citations for institutions and the top $29$\% of the articles hold about $71$\% 
of the total citations for journals.

\section*{Acknowledgments}
The authors thank S. Biswas for discussions, J.-I. Inoue, S. Redner and P. Sen for useful comments.
A.C. and B.K.C. acknowledge support from B.K.C.'s J.~C.~Bose Fellowship Research 
Grant.

\newpage


\setcounter{figure}{0}
\setcounter{table}{0}

\renewcommand{\thefigure}{S\arabic{figure}}
\renewcommand{\thetable}{S\arabic{table}}


\begin{center}
\begin{large}
\textbf{Supporting Information}
\end{large}
\end{center}

\begin{table}[!htbp]
\caption{\textbf{Abbreviations for institutions.}\\
Table showing abbreviations used for academic 
institutions.}
\label{tab:abbrevi}
\begin{tabular}{|l|l|}
\hline
Abbreviation & Full Name of University / Institute \\ \hline \hline 
Bern & University of Bern \\ \hline
BHU & Banaras Hindu University \\ \hline
Bordeaux & University of Bordeaux \\ \hline
Boston & Boston University \\ \hline 
Bristol & University of Bristol \\ \hline 
Buenos Aires & University of Buenos Aires \\ \hline
Calcutta &  University of Calcutta \\ \hline
Caltech & California Institute of Technology \\ \hline 
Cambridge & University of Cambridge   \\ \hline 
Chicago & The University of Chicago \\ \hline 
Cologne & University of Cologne \\ \hline
Columbia & Columbia University \\ \hline 
Delhi & University of 	Delhi \\ \hline 
Edinburgh & The University of Edinburgh \\ \hline 
Gottingen & University of G\"{o}ttingen \\ \hline
Groningen & University of Groningen \\ \hline 
Harvard & Harvard University \\ \hline
Heidelberg  & Heidelberg University \\ \hline 
Helsinki & University of Helsinki \\ \hline 
HUJ & The Hebrew University of Jerusalem \\ \hline 
IISC &	Indian Institute of Science \\ \hline 
\end{tabular}
\begin{tabular}{|l|l|}
\hline
Abbreviation & Full Name of University / Institute \\ \hline \hline 
Kyoto & Kyoto University \\ \hline 
Landau Inst. & Landau Institute for Theoretical Physics \\ \hline 
Leiden & Leiden University \\ \hline 
Leuven & University of Leuven -- KU Leuven \\ \hline 
Madras &  University of Madras\\ \hline 
Manchester & The University of Manchester \\ \hline 
Melbourne & The University of Melbourne\\ \hline 
MIT & Massachusetts Institute of Technology\\ \hline 
Osaka & Osaka University \\ \hline 
Oslo & University of Oslo \\ \hline 
Oxford &  University of Oxford\\ \hline 
Princeton & Princeton University \\ \hline 
SINP &	Saha Institute of Nuclear Physics \\ \hline 
Stanford & Stanford University\\ \hline 
Stockholm & Stockholm University \\ \hline 
TAU &Tel Aviv University \\ \hline 
TIFR &	Tata Institute of Fundamental Research \\ \hline 
Tokyo & The University of Tokyo  \\ \hline 
Toronto & University of Toronto\\ \hline 
Yale & Yale University\\ \hline 
Zurich & University of Zurich \\ \hline
\end{tabular}
\end{table}

\begin{figure*}[h]
\includegraphics[width=15.0cm]{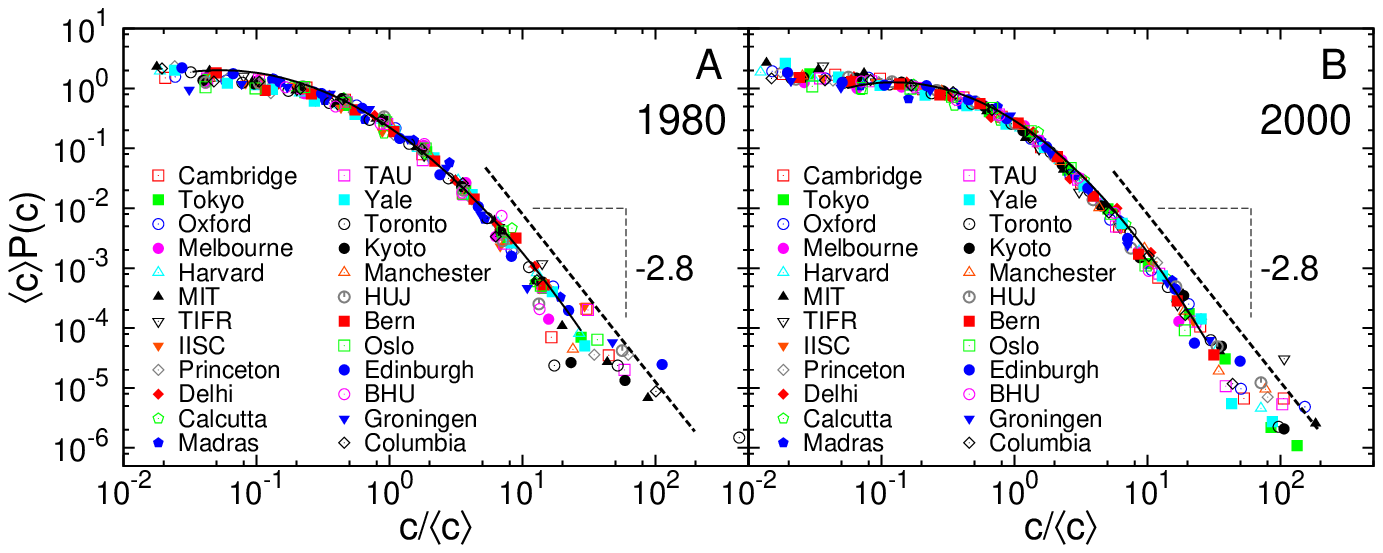}
\caption{\textbf{Probability distribution of citations for academic 
institutions for 1980, 2000.}\\
Probability distribution $P(c)$ of citations $c$ 
rescaled by average number of citations $\langle c \rangle$
 to publications from 2 different years (1980, 2000) for several academic 
institutions.
 Most of the range of the data fit well to a lognormal function with $\mu = 
-0.98  \pm 0.02$, $\sigma = 1.40 \pm 0.03$ for 1980
 and $\mu = -0.61 \pm 0.02$, $\sigma = 1.21 \pm 0.02$ for 2000.
 but the highest citations fit to power laws, with $\alpha=2.8 \pm 0.2$ for 
1980 
and $\alpha=2.8 \pm 0.1$ for 2000.
}
\label{fig:inst_80_00}
\end{figure*}

\begin{table}[!htbp]
\caption{\textbf{Abbreviation of journals, impact factors.}\\
Table showing abbreviations for journals, and 
their 2010 Impact factor~\cite{JCR}.}
\label{tab:abbrevj}
\begin{tabular}{|l|l|l|}
\hline
Abbreviation & Full Name of Journal & 2010  \\
             &                      & Impact Factor \\\hline \hline \hline
AA& Astronomy \& Astrophysics & 4.425 \\ \hline
AJ & The Astrophysical Journal & 6.063\\ \hline 
Biochemistry &	Biochemistry & 3.226 \\ \hline
CPL &	Chemical Physics Letters & 2.282 \\ \hline
Eur. J. Biochem./ & European Journal of Biochemistry (before 2005) & 3.129 \\
FEBS Journal & FEBS Journal (2005 onwards) & \\\hline 
Inor. Chem. & Inorganic Chemistry & 4.326 \\ \hline 
JAP & Journal of Applied Physics & 2.079 \\ \hline 
JCP & Journal of Chemical Physics & 2.921 \\ \hline 
JMMM &	Journal of Magnetism and Magnetic Materials & 1.690 \\ \hline 
J. Org. Chem.& Journal of Organic Chemistry &  4.002 \\ \hline 
JPA & Journal of Physics A: Mathematical and General (before 2007)  & 1.641 \\
    & Journal of Physics A: Mathematical and Theoretical (2007 onwards) & 
\\\hline 
Langmuir & Langmuir  & 4.269 \\ \hline 
Macromol. & Macromolecules  & 4.838 \\ \hline 
Physica A & Physica A  & 1.522 \\ \hline 
Physica B  & Physica B  & 0.856 \\ \hline 
Physica C & Physica C  & 1.415 \\ \hline 
PRA & Physical Review A  & 2.861 \\ \hline 
PRB &  Physical Review B  & 3.774 \\ \hline 
PRC &  Physical Review C  & 3.416 \\ \hline 
PRD &  Physical Review D  & 4.964 \\ \hline 
PRE &  Physical Review E  & 2.352 \\ \hline 
PRL &  Physical Review Letters  & 7.622 \\ \hline 
Tetrahedron & Tetrahedron  & 3.011 \\ \hline \hline 
BMJ & British Medical Journal  & 13.471\\ \hline 
Circulation & Circulation  & 14.432 \\ \hline 
JAMA &	The Journal of the American Medical Association  & 30.011 \\ \hline 
Lancet & Lancet  & 33.633 \\ \hline 
Nature & Nature  & 36.104 \\ \hline 
NEJM & The New England Journal of Medicine  & 53.486 \\ \hline
Science & Science  & 31.377 \\ \hline 
\end{tabular}
\end{table}

\begin{figure*}[h]
\includegraphics[width=10.0cm]{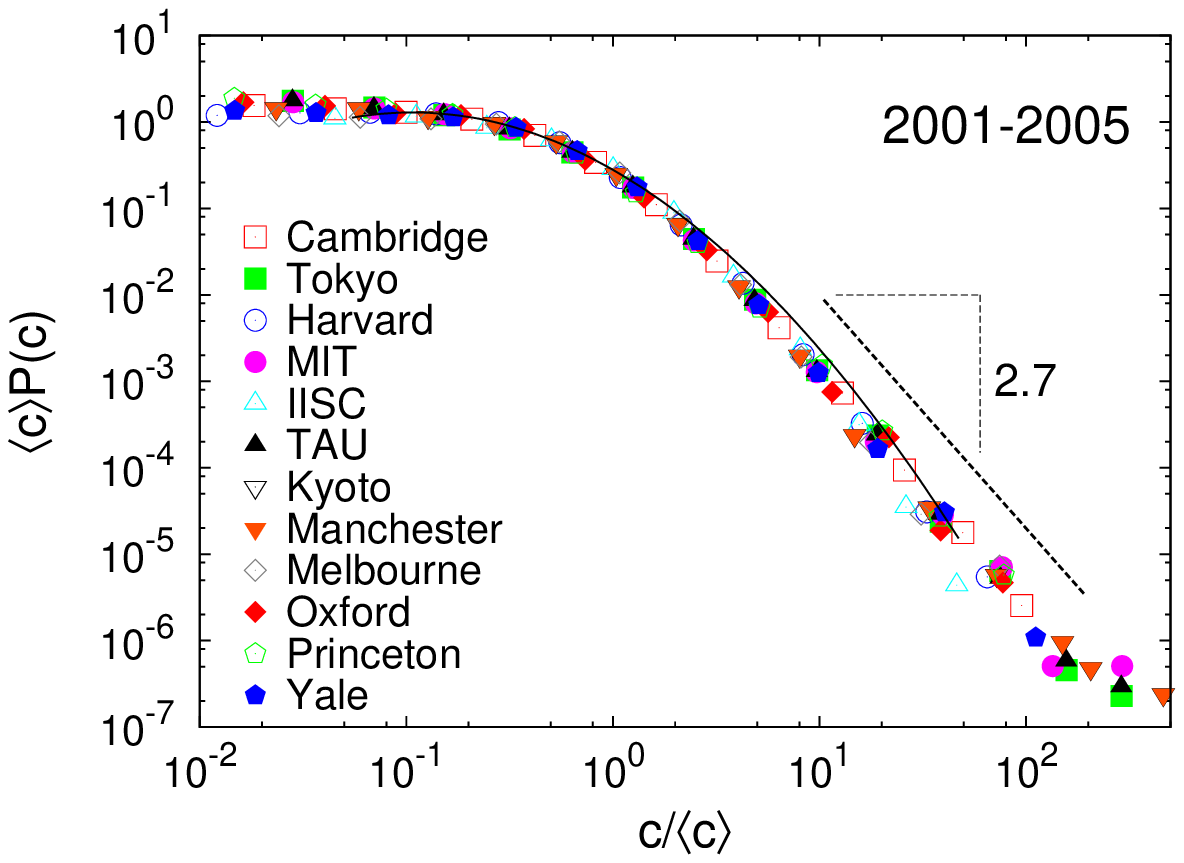}
\caption{\textbf{Probability distribution of citations for academic 
institutions, 2001-2005}\\
Probability distribution $P(c)$ of citations $c$ rescaled by average number of 
citations $\langle c \rangle$
 to publications from the period 2001-2005 for several academic institutions.
 Most of the range of the data fit well to a lognormal function with $\mu = 
-0.60 \pm 0.04$, $\sigma = 1.28 \pm 0.02$.
 but the highest citations fit to power laws, with $\alpha=2.7 \pm 0.3$.
}
\label{fig:inst_01_05}
\end{figure*}

\begin{table}[!htbp]\scriptsize
\caption{\textbf{Academic institutions: papers, citations etc.}\\
Table showing data for number of papers, total number of citations, average 
citation 
per paper $\langle c \rangle$, Gini ($g$) index, $k$ index,  
and the fraction of uncited papers $f_0$ from different institutions for 
several years.}
\label{tab:inst}
\begin{tabular}{|c|c|c|c|c|c|c|c|}
\hline

Institutions  & Year & Papers  & Citations & $\langle c \rangle$ & $g$ & $k$  & 
$f_0$\\ 
\hline

Bern   & 1980   & 	597 & 12029 & 20.15 &  0.705 & 0.767 & 0.162 \\
& 1990   & 	735 & 24468 & 33.29 &  0.684 & 0.759 & 0.076 \\
& 2000   & 	1145 & 47621 & 41.59 &  0.615 & 0.726 & 0.040 \\
& 2010   & 	1897 & 46576 & 24.55 &  0.621 & 0.726 & 0.034 \\ \hline 
BHU   & 1980   & 	454 & 2739 & 6.03 &  0.681 & 0.757 & 0.244 \\
& 1990   & 	411 & 3935 & 9.57 &  0.708 & 0.767 & 0.212 \\
& 2000   & 	371 & 5271 & 14.21 &  0.635 & 0.737 & 0.105 \\
& 2010   & 	862 & 12239 & 14.20 &  0.628 & 0.728 & 0.084 \\  \hline 
Bordeaux   & 1980   & 	412 & 7873 & 19.11 &  0.685 & 0.762 & 0.119 \\
& 1990   & 	671 & 14802 & 22.06 &  0.671 & 0.750 & 0.088 \\
& 2000   & 	1089 & 37154 & 34.12 &  0.647 & 0.740 & 0.055 \\
& 2010   & 	1903 & 41883 & 22.01 &  0.606 & 0.723 & 0.050 \\ \hline 
Boston   & 1980   & 	628 & 29097 & 46.33 &  0.661 & 0.749 & 0.065 \\
& 1990   & 	1068 & 64787 & 60.66 &  0.690 & 0.759 & 0.050 \\
& 2000   & 	1753 & 107119 & 61.11 &  0.657 & 0.744 & 0.036 \\
& 2010   & 	2829 & 89245 & 31.55 &  0.620 & 0.729 & 0.031 \\  \hline 
Bristol   & 1980   & 	595 & 22621 & 38.02 &  0.599 & 0.717 & 0.054 \\
& 1990   & 	936 & 35370 & 37.79 &  0.642 & 0.737 & 0.063 \\
& 2000   & 	1771 & 84466 & 47.69 &  0.637 & 0.736 & 0.041 \\
& 2010   & 	2479 & 63242 & 25.51 &  0.607 & 0.722 & 0.033 \\  \hline 
Buenos Aires   & 1980   & 	215 & 3392 & 15.78 &  0.804 & 0.802 & 0.293 \\
& 1990   & 	515 & 7071 & 13.73 &  0.634 & 0.734 & 0.148 \\
& 2000   & 	1100 & 28515 & 25.92 &  0.654 & 0.741 & 0.069 \\
& 2010   & 	1618 & 27120 & 16.76 &  0.669 & 0.746 & 0.069 \\ \hline 
Calcutta   & 1980   & 	146 & 794 & 5.44 &  0.697 & 0.768 & 0.219 \\
& 1990   & 	207 & 1655 & 8.00 &  0.626 & 0.735 & 0.208 \\
& 2000   & 	166 & 2365 & 14.25 &  0.650 & 0.738 & 0.157 \\
& 2010   & 	419 & 4246 & 10.13 &  0.571 & 0.710 & 0.095 \\  \hline 
Caltech   & 1980   & 	1224 & 75433 & 61.63 &  0.671 & 0.751 & 0.047 \\
& 1990   & 	1522 & 85480 & 56.16 &  0.654 & 0.743 & 0.046 \\
& 2000   & 	2228 & 153289 & 68.80 &  0.650 & 0.742 & 0.034 \\
& 2010   & 	2951 & 115107 & 39.01 &  0.642 & 0.737 & 0.028 \\  \hline 
Cambridge   & 1980   & 	1353 & 65699 & 48.56 &  0.697 & 0.762 & 0.064 \\
& 1990   & 	2273 & 118113 & 51.96 &  0.715 & 0.770 & 0.077 \\
& 2000   & 	4062 & 224825 & 55.35 &  0.674 & 0.752 & 0.048 \\
& 2010   & 	5303 & 182323 & 34.38 &  0.651 & 0.741 & 0.034 \\  \hline 
Chicago   & 1980   & 	1307 & 63672 & 48.72 &  0.646 & 0.742 & 0.047 \\
& 1990   & 	1586 & 91471 & 57.67 &  0.656 & 0.746 & 0.055 \\
& 2000   & 	2045 & 134786 & 65.91 &  0.664 & 0.747 & 0.039 \\
& 2010   & 	3285 & 114632 & 34.90 &  0.666 & 0.747 & 0.031 \\  \hline 
Cologne   & 1980   & 	586 & 11663 & 19.90 &  0.680 & 0.755 & 0.123 \\
& 1990   & 	751 & 20820 & 27.72 &  0.712 & 0.768 & 0.095 \\
& 2000   & 	1298 & 46015 & 35.45 &  0.644 & 0.740 & 0.060 \\
& 2010   & 	1727 & 37389 & 21.65 &  0.607 & 0.723 & 0.050 \\  \hline 
Columbia   & 1980   & 	1432 & 73600 & 51.40 &  0.661 & 0.747 & 0.051 \\
& 1990   & 	2021 & 123108 & 60.91 &  0.655 & 0.743 & 0.050 \\
& 2000   & 	2796 & 188207 & 67.31 &  0.644 & 0.741 & 0.032 \\
& 2010   & 	4906 & 155992 & 31.80 &  0.632 & 0.733 & 0.031 \\  \hline 
Delhi   & 1980   & 	397 & 2710 & 6.83 &  0.645 & 0.741 & 0.141 \\ 
& 1990   & 	238 & 2486 & 10.45 &  0.675 & 0.757 & 0.185 \\
& 2000   & 	278 & 4499 & 16.18 &  0.667 & 0.749 & 0.104 \\
& 2010   & 	835 & 11344 & 13.59 &  0.615 & 0.724 & 0.096 \\ \hline 
Edinburgh   & 1980   & 	723 & 26317 & 36.40 &  0.721 & 0.772 & 0.069 \\
& 1990   & 	1095 & 42946 & 39.22 &  0.654 & 0.742 & 0.070 \\
& 2000   & 	1783 & 91050 & 51.07 &  0.652 & 0.742 & 0.044 \\
& 2010   & 	2959 & 93173 & 31.49 &  0.648 & 0.740 & 0.031 \\  \hline 
Gottingen   & 1980   & 	728 & 13058 & 17.94 &  0.644 & 0.739 & 0.109 \\
& 1990   & 	966 & 38997 & 40.37 &  0.807 & 0.811 & 0.087 \\
& 2000   & 	1442 & 55161 & 38.25 &  0.657 & 0.744 & 0.057 \\
& 2010   & 	1993 & 48576 & 24.37 &  0.633 & 0.734 & 0.043 \\  \hline 
Groningen   & 1980   & 	537 & 17216 & 32.06 &  0.620 & 0.730 & 0.054 \\
& 1990   & 	937 & 36115 & 38.54 &  0.642 & 0.737 & 0.037 \\
& 2000   & 	1472 & 70664 & 48.01 &  0.612 & 0.725 & 0.026 \\
& 2010   & 	2992 & 82907 & 27.71 &  0.590 & 0.717 & 0.019 \\  \hline 
Harvard   & 1980   & 	4517 & 240905 & 53.33 &  0.695 & 0.761 & 0.144 \\
& 1990   & 	6150 & 440076 & 71.56 &  0.712 & 0.769 & 0.112 \\
& 2000   & 	8732 & 710598 & 81.38 &  0.667 & 0.750 & 0.067 \\
& 2010   & 	13446 & 474280 & 35.27 &  0.641 & 0.738 & 0.057 \\  \hline 
Heidelberg   & 1980   & 	6240 & 239840 & 38.44 &  0.638 & 0.738 & 0.029 
\\
& 1990   & 	5630 & 215294 & 38.24 &  0.640 & 0.739 & 0.029 \\
& 2000   & 	4709 & 181175 & 38.47 &  0.645 & 0.741 & 0.031 \\
& 2010   & 	3200 & 122355 & 38.24 &  0.647 & 0.742 & 0.035 \\  \hline 
Helsinki   & 1980   & 	6990 & 266778 & 38.17 &  0.634 & 0.736 & 0.029 \\
& 1990   & 	9090 & 352316 & 38.76 &  0.627 & 0.733 & 0.028 \\
& 2000   & 	10540 & 411872 & 39.08 &  0.626 & 0.733 & 0.027 \\
& 2010   & 	11106 & 429641 & 38.69 &  0.626 & 0.733 & 0.027 \\  \hline 
HUJ  & 1980   & 	1126 & 27316 & 24.26 &  0.633 & 0.732 & 0.052\\
& 1990   & 	1220 & 39586 & 32.45 &  0.639 & 0.739 & 0.050 \\
& 2000   & 	1760 & 77577 & 44.08 &  0.656 & 0.743 & 0.039 \\
& 2010   & 	1742 & 36882 & 21.17 &  0.594 & 0.717 & 0.041 \\  \hline 
IISC   & 1980   & 	406 & 4908 & 12.09 &  0.715 & 0.771 & 0.195 \\
& 1990   & 	543 & 9045 & 16.66 &  0.699 & 0.764 & 0.151 \\
& 2000   & 	826 & 22600 & 27.36 &  0.657 & 0.746 & 0.074 \\
& 2010   & 	1537 & 19963 & 12.99 &  0.606 & 0.722 & 0.076 \\  \hline

\end{tabular}
\begin{tabular}{|c|c|c|c|c|c|c|c|}
\hline

Institutions  & Year & Papers  & Citations & $\langle c \rangle$ & $g$ & $k$  & 
$f_0$\\ 
\hline

Kyoto   & 1980   & 	1861 & 47153 & 25.34 &  0.651 & 0.743 & 0.074 \\
& 1990   & 	2816 & 84563 & 30.03 &  0.662 & 0.747 & 0.063 \\
& 2000   & 	4541 & 175219 & 38.59 &  0.668 & 0.749 & 0.051 \\
& 2010   & 	5283 & 90834 & 17.19 &  0.619 & 0.728 & 0.065 \\  \hline 
Landau   & 1980   & 	94 & 8136 & 86.55 &  0.862 & 0.859 & 0.064 \\
& 1990   & 	124 & 3856 & 31.10 &  0.745 & 0.797 & 0.169 \\
& 2000   & 	232 & 5961 & 25.69 &  0.790 & 0.806 & 0.168 \\
& 2010   & 	256 & 2546 & 9.95 &  0.619 & 0.733 & 0.152 \\  \hline 
Leiden   & 1980   & 	2100 & 33516 & 15.96 &  0.612 & 0.726 & 0.075 \\
& 1990   & 	2550 & 43562 & 17.08 &  0.617 & 0.727 & 0.067 \\
& 2000   & 	2950 & 50028 & 16.96 &  0.616 & 0.727 & 0.069 \\
& 2010   & 	3650 & 63353 & 17.36 &  0.617 & 0.727 & 0.064 \\  \hline 
Leuven   & 1980   & 	2816 & 84563 & 30.03 &  0.662 & 0.747 & 0.063 \\
& 1990   & 	900 & 29257 & 32.51 &  0.692 & 0.761 & 0.062 \\
& 2000   & 	3600 & 148209 & 41.17 &  0.654 & 0.744 & 0.039 \\
& 2010   & 	950 & 38051 & 40.05 &  0.638 & 0.737 & 0.033 \\  \hline 
Madras   & 1980   & 	180 & 1363 & 7.57 &  0.666 & 0.754 & 0.183 \\
& 1990   & 	150 & 1165 & 7.77 &  0.666 & 0.756 & 0.180 \\
& 2000   & 	181 & 2608 & 14.41 &  0.622 & 0.728 & 0.144 \\
& 2010   & 	318 & 3969 & 12.48 &  0.753 & 0.783 & 0.176 \\  \hline 
Manchester   & 1980   & 	150 & 1165 & 7.77 &  0.666 & 0.756 & 0.180 \\
& 1990   & 	318 & 3969 & 12.48 &  0.753 & 0.783 & 0.176 \\
& 2000   & 	800 & 20192 & 25.24 &  0.665 & 0.747 & 0.044 \\
& 2010   & 	4289 & 107033 & 24.96 &  0.622 & 0.728 & 0.033 \\ \hline 
Melbourne   & 1980   & 	1150 & 35610 & 30.97 &  0.572 & 0.710 & 0.012 \\
& 1990   & 	850 & 22241 & 26.17 &  0.595 & 0.717 & 0.028 \\
& 2000   & 	1067 & 31538 & 29.56 &  0.606 & 0.719 & 0.023 \\
& 2010   & 	4289 & 107033 & 24.96 &  0.622 & 0.728 & 0.033 \\ \hline 
MIT   & 1980   & 	2040 & 114892 & 56.32 &  0.713 & 0.769 & 0.109 \\
& 1990   & 	2957 & 184009 & 62.23 &  0.724 & 0.777 & 0.099 \\
& 2000   & 	3524 & 258609 & 73.39 &  0.716 & 0.772 & 0.062 \\
& 2010   & 	880 & 12826 & 14.57 &  0.687 & 0.759 & 0.138 \\ \hline 
Osaka   & 1980   & 	1603 & 33261 & 20.75 &  0.624 & 0.732 & 0.069 \\
& 1990   & 	2853 & 73988 & 25.93 &  0.703 & 0.764 & 0.116 \\
& 2000   & 	450 & 14717 & 32.70 &  0.646 & 0.742 & 0.051 \\
& 2010   & 	6199 & 90104 & 14.54 &  0.680 & 0.753 & 0.125 \\ \hline 
Oslo   & 1980   & 	741 & 17956 & 24.23 &  0.658 & 0.744 & 0.113 \\
& 1990   & 	883 & 23639 & 26.77 &  0.647 & 0.740 & 0.121 \\
& 2000   & 	1410 & 46806 & 33.20 &  0.603 & 0.721 & 0.065 \\
& 2010   & 	2883 & 54926 & 19.05 &  0.587 & 0.715 & 0.059 \\  \hline 
Oxford   & 1980   & 	971 & 39763 & 40.95 &  0.647 & 0.742 & 0.048 \\
& 1990   & 	1781 & 87390 & 49.07 &  0.692 & 0.761 & 0.076 \\
& 2000   & 	3351 & 222433 & 66.38 &  0.687 & 0.756 & 0.046 \\
& 2010   & 	5539 & 199266 & 35.98 &  0.665 & 0.747 & 0.034 \\  \hline 
Princeton   & 1980   & 	1122 & 46206 & 41.18 &  0.757 & 0.788 & 0.217 \\
& 1990   & 	1474 & 73535 & 49.89 &  0.743 & 0.784 & 0.196 \\
& 2000   & 	1996 & 114847 & 57.54 &  0.714 & 0.767 & 0.139 \\
& 2010   & 	2684 & 75697 & 28.20 &  0.683 & 0.753 & 0.136 \\  \hline 
SINP   & 1980   & 	31 & 180 & 5.81 &  0.670 & 0.746 & 0.194 \\
& 1990   & 	85 & 702 & 8.26 &  0.632 & 0.731 & 0.118 \\
& 2000   & 	140 & 1576 & 11.26 &  0.648 & 0.741 & 0.157 \\
& 2010   & 	229 & 2920 & 12.75 &  0.679 & 0.752 & 0.131 \\ \hline 
Stanford   & 1980   & 	2463 & 144640 & 58.73 &  0.737 & 0.780 & 0.117 \\
& 1990   & 	3559 & 199720 & 56.12 &  0.698 & 0.764 & 0.115 \\
& 2000   & 	5541 & 393605 & 71.04 &  0.734 & 0.779 & 0.082 \\
& 2010   & 	7522 & 226442 & 30.10 &  0.679 & 0.755 & 0.082 \\  \hline 
Stockholm   & 1980   & 	410 & 11393 & 27.79 &  0.695 & 0.762 & 0.124 \\
& 1990   & 	648 & 22535 & 34.78 &  0.664 & 0.752 & 0.094 \\
& 2000   & 	521 & 25832 & 49.58 &  0.685 & 0.756 & 0.060 \\
& 2010   & 	312 & 8346 & 26.75 &  0.698 & 0.762 & 0.090 \\ \hline 
TAU   & 1980   & 	1145 & 26937 & 23.53 &  0.718 & 0.770 & 0.114 \\
& 1990   & 	1761 & 40134 & 22.79 &  0.679 & 0.751 & 0.114 \\
& 2000   & 	2657 & 77106 & 29.02 &  0.663 & 0.746 & 0.074 \\
& 2010   & 	3402 & 55252 & 16.24 &  0.657 & 0.744 & 0.103 \\ \hline 
TIFR   & 1980   & 	163 & 2042 & 12.53 &  0.699 & 0.765 & 0.153 \\
& 1990   & 	322 & 5256 & 16.32 &  0.745 & 0.780 & 0.196 \\
& 2000   & 	437 & 12043 & 27.56 &  0.736 & 0.774 & 0.114 \\
& 2010   & 	578 & 13803 & 23.88 &  0.747 & 0.778 & 0.104 \\  \hline 
Tokyo   & 1980   & 	2595 & 60693 & 23.39 &  0.666 & 0.748 & 0.124 \\
& 1990   & 	4383 & 121428 & 27.70 &  0.677 & 0.754 & 0.094 \\
& 2000   & 	7734 & 267430 & 34.58 &  0.676 & 0.752 & 0.079 \\
& 2010   & 	8980 & 164182 & 18.28 &  0.655 & 0.743 & 0.085 \\  \hline 
Toronto   & 1980   & 	2568 & 80164 & 31.22 &  0.771 & 0.793 & 0.220 \\
& 1990   & 	3613 & 116662 & 32.29 &  0.714 & 0.769 & 0.184 \\
& 2000   & 	5185 & 246853 & 47.61 &  0.684 & 0.756 & 0.123 \\
& 2010   & 	8880 & 207596 & 23.38 &  0.649 & 0.739 & 0.091 \\  \hline 
Yale   & 1980   & 	2400 & 99265 & 41.36 &  0.716 & 0.770 & 0.202 \\
& 1990   & 	2940 & 165230 & 56.20 &  0.725 & 0.774 & 0.162 \\
& 2000   & 	4707 & 249190 & 52.94 &  0.723 & 0.773 & 0.163 \\
& 2010   & 	6588 & 156268 & 23.72 &  0.684 & 0.756 & 0.143 \\  \hline 
Zurich   & 1980   & 	712 & 16173 & 22.71 &  0.718 & 0.773 & 0.154 \\
& 1990   & 	844 & 28233 & 33.45 &  0.684 & 0.756 & 0.120 \\
& 2000   & 	1991 & 97451 & 48.95 &  0.661 & 0.748 & 0.065 \\
& 2010   & 	3423 & 87603 & 25.59 &  0.629 & 0.729 & 0.059 \\  \hline

\end{tabular}
\end{table}

\begin{table}\scriptsize
\caption{\textbf{Academic journals: papers, citations etc.}\\
Table showing data for number of papers, total number of citations, average 
citation 
per paper $\langle c \rangle$, Gini ($g$) index, $k$ index, 
and the fraction of uncited papers $f_0$ from different journals for 
several years.}
\label{tab:jrnlg}
\begin{tabular}{|c|c|c|c|c|c|c|c|}
\hline

Journals  & Year & papers  & citations & $\langle c \rangle$ & $g$ & $k$  & 
$f_0$\\ 
\hline

Astronomy & 1980 & 	728 & 20594 & 28.29 &  0.636 & 0.734 & 0.026\\ 
Astrophys. & 1990 & 	909 & 27208 & 29.93 &  0.577 & 0.715 & 0.019\\ 
& 2000 & 	1412 & 51354 & 36.37 &  0.558 & 0.704 & 0.018\\ 
& 2010 & 	1916 & 40226 & 20.99 &  0.564 & 0.704 & 0.027\\ \hline
Astrophys. & 1980 & 	1223 & 64099 & 52.41 &  0.550 & 0.701 & 0.008\\ 
 J.& 1990 & 	1517 & 72110 & 47.53 &  0.533 & 0.696 & 0.010\\ 
& 2000 & 	2388 & 136940 & 57.35 &  0.547 & 0.701 & 0.003\\ 
& 2010 & 	2501 & 73646 & 29.45 &  0.506 & 0.685 & 0.007\\ \hline
Biochem. & 1980 & 	935 & 55002 & 58.83 &  0.486 & 0.674 & 0.002\\ 
& 1990 & 	1510 & 104554 & 69.24 &  0.509 & 0.684 & 0.001\\ 
& 2000 & 	1823 & 79977 & 43.87 &  0.435 & 0.656 & 0.001\\ 
& 2010 & 	1142 & 18512 & 16.21 &  0.436 & 0.654 & 0.007\\ \hline
BMJ & 1980 & 	809 & 26008 & 32.15 &  0.676 & 0.757 & 0.080\\ 
& 1990 & 	626 & 31949 & 51.04 &  0.692 & 0.763 & 0.054\\ 
& 2000 & 	613 & 61513 & 100.35 &  0.709 & 0.769 & 0.038\\ 
& 2010 & 	293 & 14339 & 48.94 &  0.507 & 0.682 & 0.010\\ \hline
Circulation & 1980 & 	412 & 32786 & 79.58 &  0.555 & 0.704 & 0.022\\ 
& 1990 & 	541 & 62367 & 115.28 &  0.571 & 0.713 & 0.004\\ 
& 2000 & 	989 & 130463 & 131.91 &  0.528 & 0.693 & 0.004\\ 
& 2010 & 	554 & 41659 & 75.20 &  0.492 & 0.675 & 0.004\\ \hline
CPL & 1980 & 	1067 & 31538 & 29.56 &  0.606 & 0.719 & 0.023\\ 
& 1990 & 	1166 & 34366 & 29.47 &  0.627 & 0.730 & 0.025\\ 
& 2000 & 	1487 & 47478 & 31.93 &  0.579 & 0.713 & 0.011\\ 
& 2010 & 	1013 & 10966 & 10.83 &  0.525 & 0.687 & 0.051\\ \hline
Eur. J. & 1980 & 	753 & 35476 & 47.11 &  0.545 & 0.697 & 0.005\\  
Biochem. & 1990 & 	789 & 35456 & 44.94 &  0.531 & 0.693 & 0.009\\ 
& 2000 & 	836 & 34696 & 41.50 &  0.514 & 0.683 & 0.004\\ 
& 2010 & 	447 & 9702 & 21.70 &  0.545 & 0.698 & 0.022\\ \hline
Inor. Chem. & 1980 & 	823 & 31391 & 38.14 &  0.459 & 0.666 & 0.001\\ 
& 1990 & 	1082 & 39076 & 36.11 &  0.476 & 0.672 & 0.004\\ 
& 2000 & 	941 & 36648 & 38.95 &  0.466 & 0.668 & 0.002\\ 
& 2010 & 	1416 & 31953 & 22.57 &  0.447 & 0.662 & 0.007\\ \hline
JAMA & 1980 & 	503 & 13999 & 27.83 &  0.675 & 0.753 & 0.205\\ 
& 1990 & 	743 & 48102 & 64.74 &  0.762 & 0.787 & 0.353\\ 
& 2000 & 	699 & 79037 & 113.07 &  0.757 & 0.788 & 0.330\\ 
& 2010 & 	535 & 34563 & 64.60 &  0.723 & 0.772 & 0.187\\ \hline
JAP & 1980 & 	1108 & 34304 & 30.96 &  0.668 & 0.754 & 0.046\\ 
& 1990 & 	2571 & 57132 & 22.22 &  0.638 & 0.739 & 0.054\\ 
& 2000 & 	2941 & 80656 & 27.42 &  0.613 & 0.727 & 0.027\\ 
& 2010 & 	3892 & 40286 & 10.35 &  0.511 & 0.685 & 0.053\\ \hline
J. Chem.& 1980 & 	1849 & 101604 & 54.95 &  0.651 & 0.739 & 0.016\\ 
Phys. & 1990 & 	1958 & 95959 & 49.01 &  0.582 & 0.711 & 0.012\\ 
& 2000 & 	2526 & 102525 & 40.59 &  0.579 & 0.710 & 0.005\\ 
& 2010 & 	2137 & 32872 & 15.38 &  0.522 & 0.686 & 0.022\\ \hline
JMMM & 1980 & 	907 & 7467 & 8.23 &  0.631 & 0.735 & 0.128\\ 
& 1990 & 	852 & 8734 & 10.25 &  0.653 & 0.744 & 0.131\\ 
& 2000 & 	802 & 12714 & 15.85 &  0.584 & 0.714 & 0.062\\ 
& 2010 & 	740 & 6382 & 8.62 &  0.570 & 0.708 & 0.099\\ \hline
J. Org.& 1980 & 	1232 & 38706 & 31.42 &  0.513 & 0.687 & 0.004\\ 
Chem. & 1990 & 	1162 & 44254 & 38.08 &  0.494 & 0.680 & 0.001\\ 
& 2000 & 	1396 & 53495 & 38.32 &  0.442 & 0.659 & 0.002\\ 
& 2010 & 	1201 & 25166 & 20.95 &  0.417 & 0.649 & 0.005\\ \hline
JPA & 1980 & 	340 & 7800 & 22.94 &  0.752 & 0.790 & 0.106\\ 
& 1990 & 	470 & 7507 & 15.97 &  0.625 & 0.735 & 0.085\\ 
& 2000 & 	644 & 8758 & 13.60 &  0.592 & 0.722 & 0.079\\ 
& 2010 & 	959 & 6545 & 6.82 &  0.573 & 0.707 & 0.115\\ \hline
Lancet & 1980 & 	595 & 38780 & 65.18 &  0.650 & 0.736 & 0.229\\ 
& 1990 & 	476 & 67331 & 141.45 &  0.604 & 0.721 & 0.048\\ 
& 2000 & 	822 & 109177 & 132.82 &  0.642 & 0.739 & 0.050\\ 
& 2010 & 	271 & 54158 & 199.85 &  0.463 & 0.670 & 0.000\\ \hline

\end{tabular}
\begin{tabular}{|c|c|c|c|c|c|c|c|}
\hline

Journals  & Year & papers  & citations & $\langle c \rangle$ & $g$ & $k$  & 
$f_0$\\ 
\hline

Langmuir & 1990 & 	285 & 12679 & 44.49 &  0.589 & 0.716 & 0.014\\ 
& 2000 & 	1476 & 72398 & 49.05 &  0.528 & 0.694 & 0.003\\ 
& 2010 & 	2664 & 55540 & 20.85 &  0.460 & 0.665 & 0.007\\ \hline
Macromol. & 1980 & 	315 & 20018 & 63.55 &  0.642 & 0.737 & 0.029\\ 
& 1990 & 	863 & 42799 & 49.59 &  0.567 & 0.710 & 0.008\\ 
& 2000 & 	1373 & 69158 & 50.37 &  0.499 & 0.682 & 0.001\\ 
& 2010 & 	1365 & 38215 & 28.00 &  0.472 & 0.668 & 0.006\\ \hline
Nature & 1980 & 	1502 & 181108 & 120.58 &  0.637 & 0.736 & 0.007\\ 
& 1990 & 	1391 & 315723 & 226.98 &  0.676 & 0.751 & 0.180\\ 
& 2000 & 	1517 & 464531 & 306.22 &  0.668 & 0.746 & 0.146\\ 
& 2010 & 	1012 & 163098 & 161.16 &  0.547 & 0.698 & 0.111\\ \hline
NEJM & 1980 & 	360 & 67263 & 186.84 &  0.518 & 0.686 & 0.003\\ 
& 1990 & 	374 & 111199 & 297.32 &  0.506 & 0.683 & 0.000\\ 
& 2000 & 	379 & 142799 & 376.78 &  0.603 & 0.720 & 0.055\\ 
& 2010 & 	342 & 77333 & 226.12 &  0.576 & 0.706 & 0.029\\ \hline
Physica A & 1980 & 	195 & 3298 & 16.91 &  0.551 & 0.700 & 0.067\\ 
& 1990 & 	402 & 6691 & 16.64 &  0.653 & 0.748 & 0.112\\ 
& 2000 & 	620 & 10776 & 17.38 &  0.649 & 0.744 & 0.077\\ 
& 2010 & 	617 & 5217 & 8.46 &  0.587 & 0.718 & 0.100\\ \hline
Physica B & 1990 & 	1187 & 6787 & 5.72 &  0.632 & 0.732 & 0.190\\ 
& 2000 & 	2630 & 13378 & 5.09 &  0.647 & 0.740 & 0.220\\ 
& 2010 & 	1058 & 7139 & 6.75 &  0.558 & 0.702 & 0.127\\ \hline
Physica C & 1990 & 	608 & 16779 & 27.60 &  0.586 & 0.715 & 0.016\\ 
& 2000 & 	1621 & 8481 & 5.23 &  0.664 & 0.748 & 0.240\\ 
& 2010 & 	897 & 2528 & 2.82 &  0.658 & 0.744 & 0.309\\ \hline
PRA & 1980 & 	624 & 25452 & 40.79 &  0.609 & 0.724 & 0.024\\ 
& 1990 & 	1859 & 57309 & 30.83 &  0.603 & 0.724 & 0.033\\ 
& 2000 & 	1410 & 42545 & 30.17 &  0.624 & 0.729 & 0.044\\ 
& 2010 & 	2858 & 39628 & 13.87 &  0.519 & 0.687 & 0.031\\ \hline
PRB & 1980 & 	1354 & 64235 & 47.44 &  0.648 & 0.743 & 0.018\\ 
& 1990 & 	3390 & 161651 & 47.68 &  0.649 & 0.741 & 0.017\\ 
& 2000 & 	4756 & 172663 & 36.30 &  0.602 & 0.722 & 0.025\\ 
& 2010 & 	6049 & 108383 & 17.92 &  0.528 & 0.692 & 0.021\\ \hline
PRC & 1980 & 	619 & 19945 & 32.22 &  0.653 & 0.744 & 0.027\\ 
& 1990 & 	703 & 16018 & 22.79 &  0.617 & 0.728 & 0.033\\ 
& 2000 & 	832 & 20521 & 24.66 &  0.569 & 0.709 & 0.040\\ 
& 2010 & 	1013 & 18071 & 17.84 &  0.545 & 0.697 & 0.032\\ \hline
PRD & 1980 & 	775 & 37491 & 48.38 &  0.763 & 0.797 & 0.081\\ 
& 1990 & 	1007 & 34899 & 34.66 &  0.681 & 0.759 & 0.049\\ 
& 2000 & 	2020 & 70517 & 34.91 &  0.613 & 0.728 & 0.040\\ 
& 2010 & 	2931 & 56035 & 19.12 &  0.532 & 0.693 & 0.028\\ \hline
PRE & 2000 & 	2034 & 58313 & 28.67 &  0.584 & 0.715 & 0.022\\ 
& 2010 & 	2310 & 26163 & 11.33 &  0.492 & 0.678 & 0.039\\ \hline
PRL & 1980 & 	1194 & 92238 & 77.25 &  0.670 & 0.746 & 0.047\\ 
& 1990 & 	1643 & 163084 & 99.26 &  0.604 & 0.724 & 0.003\\ 
& 2000 & 	3046 & 250871 & 82.36 &  0.589 & 0.717 & 0.006\\ 
& 2010 & 	3105 & 115955 & 37.34 &  0.493 & 0.679 & 0.004\\ \hline
Science & 1980 & 	1021 & 115716 & 113.34 &  0.635 & 0.738 & 0.032\\ 
& 1990 & 	1061 & 237803 & 224.13 &  0.663 & 0.745 & 0.168\\ 
& 2000 & 	1053 & 343455 & 326.17 &  0.614 & 0.725 & 0.092\\ 
& 2010 & 	974 & 135833 & 139.46 &  0.529 & 0.692 & 0.089\\ \hline
Tetrahedron & 1980 & 	391 & 16696 & 42.70 &  0.709 & 0.771 & 0.013\\ 
& 1990 & 	684 & 19410 & 28.38 &  0.556 & 0.701 & 0.034\\ 
& 2000 & 	1092 & 33027 & 30.24 &  0.503 & 0.680 & 0.007\\ 
& 2010 & 	1179 & 16155 & 13.70 &  0.465 & 0.665 & 0.015\\ \hline
&&&&&&&\\
&&&&&&&\\
&&&&&&&\\
&&&&&&&\\
&&&&&&&\\\hline

\end{tabular}
\end{table}
 
\begin{table}
\caption{\textbf{Elite journals, considering all citable documents: papers, 
citations etc.}\\
Table showing data 
for number of citable documents, total number of citations, average citation 
per paper $\langle c \rangle$, Gini ($g$) index, $k$ index, 
as well as the fraction of uncited papers $f_0$ from different journals of the 
\textit{Elite} class for several years.}
\label{tab:jrnle}
\begin{tabular}{|c|c|c|c|c|c|c|c|}
\hline

Journals  & Year & papers  & citations & $\langle c \rangle$ & $g$ & $k$  & 
$f_0$\\ 
\hline

BMJ &	1980	& 3056 & 27798 & 9.10 &  0.886 & 0.869 & 0.542 \\
 &	1990	& 2824 & 37353 & 13.23 &  0.884 & 0.867 & 0.482 \\
 &	2000	& 3263 & 71373 & 21.87 &  0.901 & 0.877 & 0.451 \\
 &	2010	& 3412 & 19308 & 5.66 &  0.900 & 0.883 & 0.601 \\ \hline

Circulation &	1980	& 1895 & 35986 & 18.99 &  0.861 & 0.861 & 0.387 \\
 &	1990	& 3830 & 70278 & 18.35 &  0.907 & 0.891 & 0.478 \\
 &	2000	& 5574 & 137302 & 24.63 &  0.906 & 0.887 & 0.628 \\
 &	2010	& 6675 & 46645 & 6.99 &  0.951 & 0.931 & 0.827 \\ \hline 

JAMA &	1980	& 1440 & 20784 & 14.43 &  0.794 & 0.812 & 0.386 \\
 &	1990	& 1738 & 54443 & 31.33 &  0.876 & 0.862 & 0.472 \\
 &	2000	& 1696 & 86713 & 51.13 &  0.869 & 0.862 & 0.381 \\
 &	2010	& 1425 & 36235 & 25.43 &  0.863 & 0.857 & 0.435 \\ \hline 

Lancet &	1980	& 2981 & 56515 & 18.96 &  0.822 & 0.829 & 0.378 \\
 &	1990	& 3230 & 94008 & 29.10 &  0.846 & 0.842 & 0.344 \\
 &	2000	& 3367 & 123585 & 36.70 &  0.870 & 0.863 & 0.365 \\
 &	2010	& 1743 & 62028 & 35.59 &  0.876 & 0.873 & 0.368 \\ \hline 

Nature &	1980	& 2892 & 185484 & 64.14 &  0.799 & 0.809 & 0.336  \\
 &	1990	& 3606 & 318090 & 88.22 &  0.857 & 0.847 & 0.435 \\
 &	2000	& 3612 & 330512 & 91.50 &  0.856 & 0.847 & 0.434 \\
 &	2010	&  2577 & 177161 & 68.75 &  0.791 & 0.809 & 0.302 \\ \hline

NEJM &	1980	& 1791 & 77780 & 43.43 &  0.858 & 0.855 & 0.376 \\
 &	1990	& 1684 & 122750 & 72.89 &  0.854 & 0.851 & 0.348 \\
 &	2000	& 1561 & 155490 & 99.61 &  0.874 & 0.864 & 0.336 \\
 &	2010	& 1753 & 93609 & 53.40 &  0.867 & 0.861 & 0.345 \\ \hline
 
Science &	1980	& 1669 & 117642 & 70.49 &  0.765 & 0.795 & 0.217\\
 &	1990	& 2178 & 243190 & 111.66 &  0.826 & 0.829 & 0.354\\
 &	2000	& 2575 & 363418 & 141.13 &  0.816 & 0.823 & 0.260 \\
 &	2010	& 2439 & 154194 & 63.22 &  0.762 & 0.795 & 0.243 \\ \hline 

\end{tabular}
\end{table}

\begin{figure*}
\includegraphics[width=17.5cm]{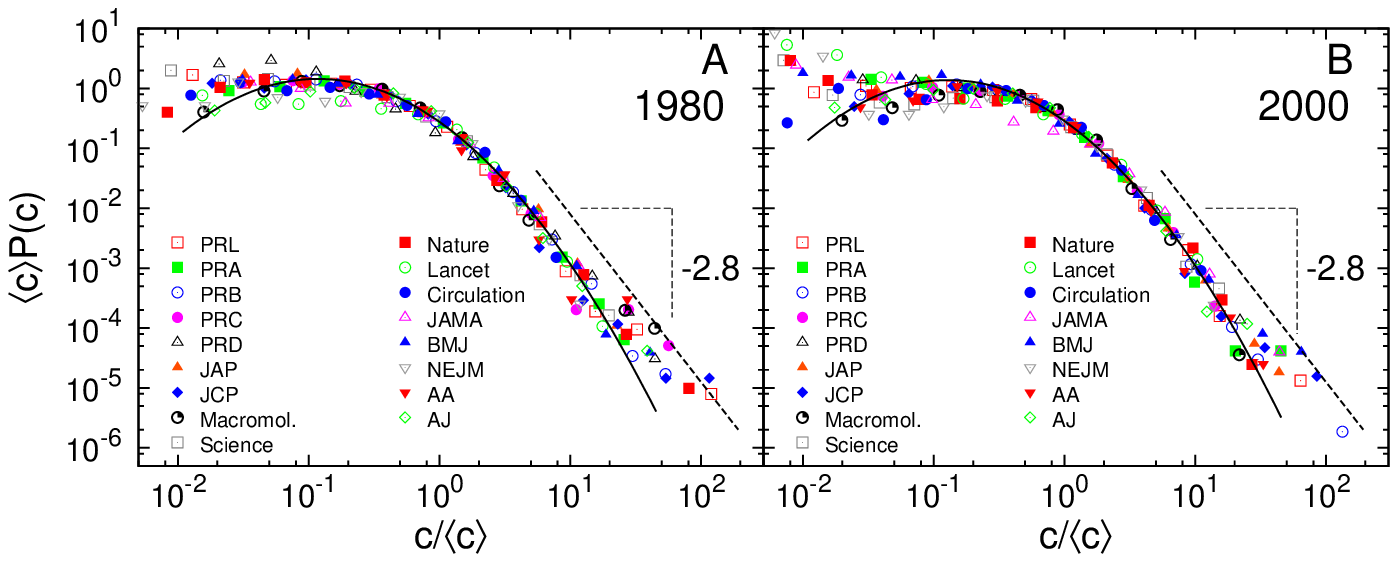}
\caption{\textbf{Probability distribution of citations for journals for 1980, 
2000.}\\
Probability distribution $P(c)$ of citations $c$ rescaled by average number of 
citations $\langle c \rangle$
 to publications from 2 different years, (A) 1980 and (B) 2000 for several 
journals.
 The most of the range fits to a lognormal with 
$\mu = -0.75 \pm 0.02$, $\sigma = 1.18 \pm 0.02$ for 1980
 and $\mu = -0.72 \pm 0.02$, $\sigma = 1.15 \pm 0.03$ for 2000.
  The largest citations for both the classes seem to follow a power law: 
$c^{-\alpha}$, with $\alpha = 2.8 \pm 0.4$
 for 1980 and $\alpha = 2.8 \pm 0.3$ for 2000.
}
\label{fig:jour_80_00}
\end{figure*}
\begin{figure*}
\includegraphics[width=10.0cm]{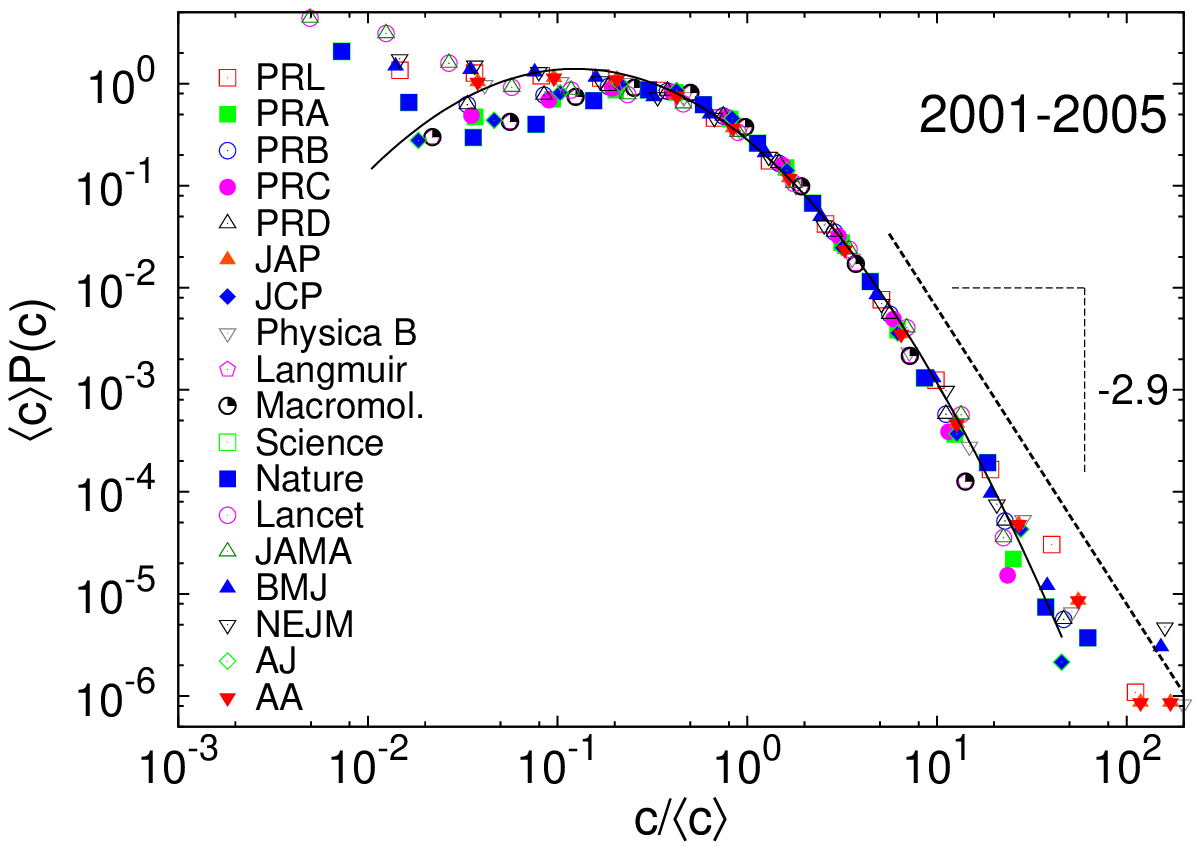}
\caption{\textbf{Probability distribution of citations for journals for 
2001-2005.}\\
Probability distribution $P(c)$ of citations $c$ rescaled by average number of 
citations $\langle c \rangle$
 to publications from 2001-2005 for several journals.
 The most of the range fits to a lognormal with 
$\mu = -0.73 \pm 0.02$, $\sigma = 1.16 \pm 0.02$, while the largest citations 
seem to follow a power law: $c^{-\alpha}$, with $\alpha = 
2.9 \pm 0.2$.
}
\label{fig:joun_01_05}
\end{figure*}

\begin{figure*}[h]
\includegraphics[width=10.0cm]{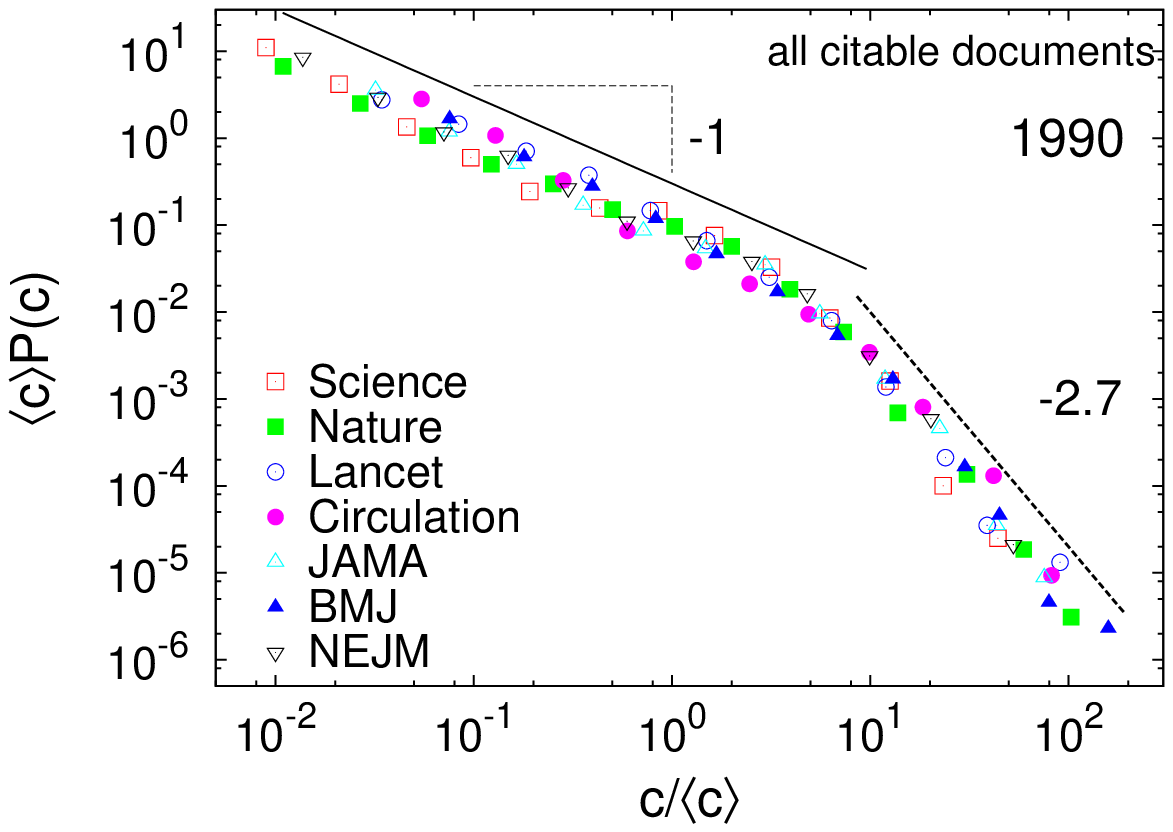}
\caption{\textbf{Rescaled probability distributions of citation to all citable 
documents for several journals of the Elite class for 1990}\\
Probability distribution $P(c)$ of citations $c$ rescaled by average number of 
citations $\langle c \rangle$
to publications from  1990 for
all citable documents of several academic journals in the \textit{Elite} class: 
The scaling distribution is such that
$\langle c \rangle P(c) \sim (c/\langle c \rangle)^{-b}$ with $b \simeq 1$,
for the lower range of $c$.
The largest citations fit well to a power law: $c^{-\alpha}$, with $\alpha = 
2.7 \pm 0.4$.
}
\label{fig:joun2_elite}
\end{figure*}

\begin{figure*}[h]
\includegraphics[width=17.0cm]{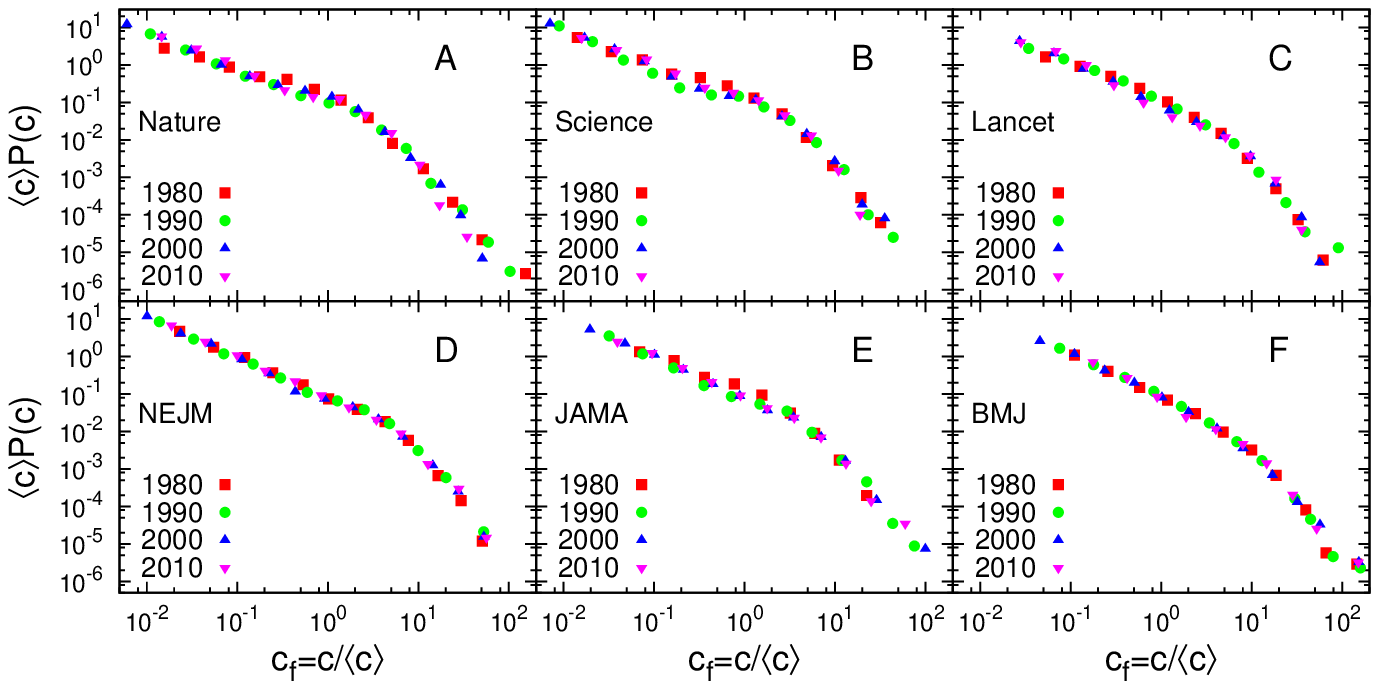}
\caption{\textbf{Rescaled probability distributions of citation to all citable 
documents for several journals of the Elite class for different 
years.}\\
Probability distribution $P(c)$ of citations $c$ rescaled by average number of 
citations $\langle c \rangle$
to publications from  1980, 1990, 2000 and 2010 for
all citable documents of several academic journals in the \textit{Elite} class.
}
\label{fig:joun_elite}
\end{figure*}

\begin{figure*}
\includegraphics[width=17.5cm]{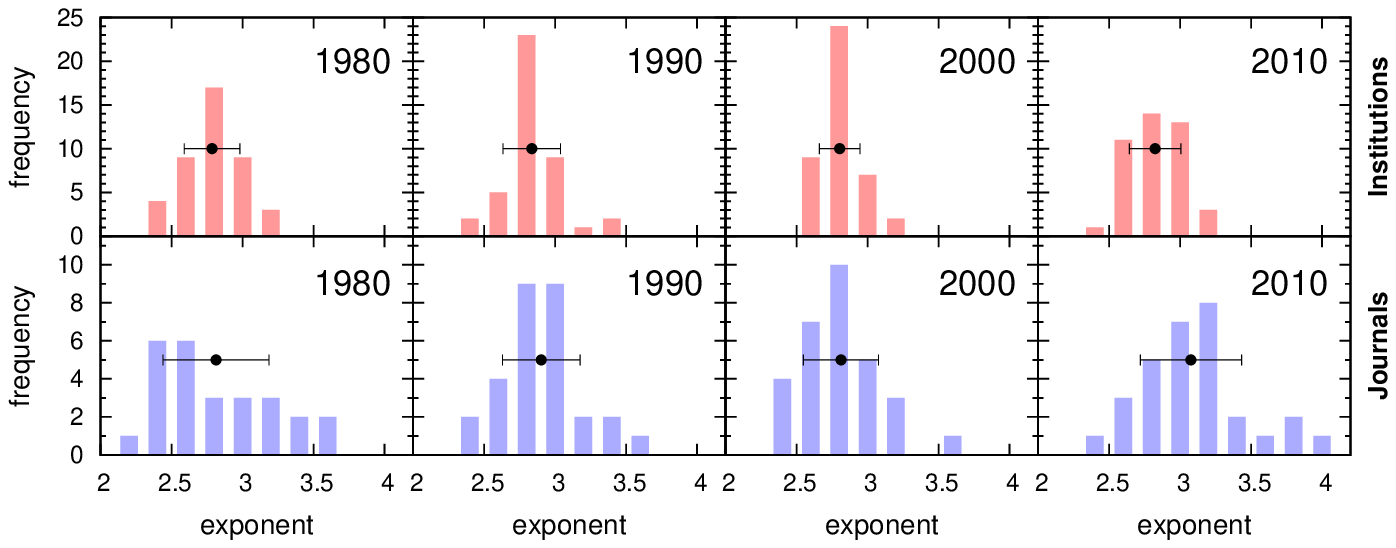}
\caption{\textbf{Histograms for power law exponents.}\\
Histograms of power law exponents of the highest cited papers measured from 
individual data sets, their average along with error bar.
Data is shown for 4 different years: 1980, 1990, 2000 and 2010,
 for institutions and journals.
}
\label{fig:exp}
\end{figure*}

\begin{table}[h]
\caption{\textbf{Power law exponents.}\\
Average power law exponents for the highest cited publications for different 
years (1980, 1990, 2000 and 2010), for all institutions and journals 
considered.}
\label{tab:exp}
\begin{tabular}{|l|c|c|}
\hline
Year  & Institutions & Journals\\ 
\hline
1980 &  $2.8 \pm 0.2$ & $2.8 \pm 0.4$ \\ \hline
1990 &  $2.8 \pm 0.2$  &  $2.9 \pm  0.3$  \\ \hline 
2000 & $2.8 \pm 0.1$  & $2.8 \pm  0.3$ \\ \hline 
2010 & $2.8 \pm 0.2$  & $3.0 \pm 0.4$  \\ \hline
\end{tabular}
\end{table}

\end{document}